\documentclass[%
 aip,
 cha,
 amsmath,amssymb,
 floatfix,
 reprint,%
]{revtex4-1}
\usepackage{dcolumn}
\usepackage{bm}

\usepackage[utf8]{inputenc}
\usepackage[T1]{fontenc}
\usepackage{mathptmx}
\usepackage{etoolbox}
\usepackage{graphicx} 
\usepackage{hyperref}
\hypersetup{
    colorlinks=true,
    linkcolor=blue,
    urlcolor=blue,
}
 \usepackage{amsmath}
\usepackage{amsfonts}
\usepackage{booktabs}

\usepackage{algorithmic}
\usepackage{amssymb}
\usepackage{lipsum}
\usepackage{epstopdf}
\usepackage{geometry}
\usepackage{dcolumn}
\usepackage{bm}
\usepackage[utf8]{inputenc}
\usepackage[T1]{fontenc}
\usepackage{mathptmx}
\usepackage{etoolbox}
\usepackage{xcolor}

\newcommand{\PIK}{Potsdam Institute for Climate Impact Research, Telegrafenberg, 14473 Potsdam, Germany}
\newcommand{\UCC}{School of Mathematical Sciences, University College Cork, Cork, Ireland}
\newcommand{\IM}{Institute of Mathematics, National Academy of Sciences of Ukraine, 
01024 Kyiv, Ukraine}
\newcommand{\HU}{Institute of Mathematics, Humboldt University Berlin, 12489 Berlin, Germany}
\newcommand{\FUDAN}{Fudan University, Institute Intelligent Complex Systems, Shanghai, China}
\newcommand{\JIANGSU}{Faculty of Civil Engineering and Mechanics, Jiangsu University, Zhenjiang 212013, China}
\newcommand{\YANGZHOU}{School of Mathematical Science, Yangzhou University, Yangzhou 225002, China}

\preprint{AIP/123-QED}
\begin{document}


\title{Synchronization cluster bursting in adaptive oscillators networks}

\author{Mengke Wei}
\affiliation{\YANGZHOU}
\affiliation{\PIK}
\affiliation{\JIANGSU}

\author{Andreas Amann}
\affiliation{\PIK}
\affiliation{\UCC}

\author{Oleksandr Burylko}
\affiliation{\PIK}
\affiliation{\IM}
\affiliation{\HU}

\author{Xiujing Han}
\affiliation{\JIANGSU}

\author{Serhiy Yanchuk}
\affiliation{\UCC}

\author{J\"urgen Kurths}
\affiliation{\PIK}
\affiliation{\FUDAN}

\date{\today}

\begin{abstract}
Adaptive dynamical networks are ubiquitous in real-world systems. This paper aims to explore the synchronization dynamics in networks of adaptive oscillators based on a paradigmatic system of adaptively coupled phase oscillators. Our numerical observations reveal the emergence of synchronization cluster bursting, characterized by periodic transitions between cluster synchronization and global synchronization. 
By investigating a reduced model, the mechanisms underlying synchronization cluster bursting are clarified. 
We show that a minimal model exhibiting this phenomenon can be reduced to a phase oscillator with complex-valued adaptation. 
Furthermore, the adaptivity of the system leads to the appearance of additional symmetries and thus to the coexistence of stable bursting solutions with very different Kuramoto order parameters. 
\end{abstract}


\maketitle 
\begin{quotation}
Synchronization within dynamical systems on networks is frequently encountered in various fields of natural sciences and engineering technology. Recently, the study of synchronization of coupled systems has attracted extensive attention, and plenty of new patterns of network synchronization, such as complete synchronization, cluster synchronization, remote synchronization, and chimera states, have been reported. In particular, networks with adaptive couplings have emerged as a focal point of research due to their ability to model complex interactions more realistically. In this paper, we explore the synchronization dynamics of an adaptive oscillator network. Our results show that synchronization cluster bursting can be observed numerically in a paradigmatic system of adaptively coupled phase oscillators. We investigate the symmetries of the model and reduce the model to a normal form to reveal the mechanism of synchronization cluster bursting. Based on the dynamical analysis of the normal form equation, we point out the role of fixed points in synchronization state transitions and the generation of bursting.

\end{quotation}

\section{Introduction}
\label{Introduction}

The aspect of adaptivity  \cite{christensen1998evolution,bornholdt2000topological,gross2008adaptive,Berner2023,Sawicki2023} has introduced a new dimension to the classic study of coupled oscillator networks. In an adaptive network the coupling between oscillators evolves depending on the state of the system. The main motivation for studying adaptive networks stems from its importance in real world systems \cite{sayama2013modeling,Berner2023}. For instance, in biological networks such as the human brain, synaptic plasticity allows for a strengthening or weakening of connections between neurons, facilitating learning and memory \cite{hebb1949organization,deneve2017brain}. Similarly, social networks exhibit adaptability, constantly reshaping connections to improve collective intelligence and accurate individual and collective beliefs \cite{almaatouq2020adaptive}. Even in ecological networks, species adapt their behaviors in response to changes in their habitat or population dynamics, thus promoting the stability of community dynamics \cite{landi2018complexity,raimundoAdaptiveNetworksRestoration2018}. It is therefore important to understand the basic mechanisms of emergent collective behaviors and patterns in adaptive oscillator networks\cite{maistrenko2007multistability,gutierrez2011emerging,Burylko2018Winner}.

In many complex systems of coupled oscillators \cite{keane2023transitional,burylko2018coexistence,reviews3,physics3,berner2019multiclusters,FIA23} it is possible to reduce much of the complexity by focusing on the dynamics of carefully chosen phase variables. The system is thus transformed into a model of coupled phase oscillators, which allows for a simplified analysis of the fundamental properties of the system. In this paper, we consider an adaptive version of the Kuramoto phase oscillator model.
The classic Kuramoto model is an idealized model which has been successfully applied to coupled oscillator systems arising in physics \cite{physics1,physics2,physics4}, chemistry \cite{chemicaloscillators1,chemicaloscillators2}, electrical engineering \cite{electrical_engineering1,electrical_engineering2,electrical_engineering3,electrical_engineering4} and other fields. The Kuramoto model and its many variations and extensions has been extensively studied in the literature. For example, an opinion changing rate model \cite{sociophysics1} was proposed by modifying the classical Kuramoto model to investigate the opinion synchronization in social networks. The Kuramoto model with phase lag was applied to the non-linear dynamics on a directed graph of a sequence of earthquakes \cite{earthquake}. A number of recent reviews  \cite{reviews1,reviews2,reviews3,reviews4,reviews5} provide a detailed survey of the  Kuramoto model, its extensions and its most significant applications. 

Synchronization \cite{motter2005network,arenas2008synchronization,boccaletti2018synchronization}, as a collective behavior among populations of dynamically interacting entities, has been extensively studied for its significant contributions in various fields, including biology \cite{strogatz1993coupled}, ecology \cite{blasius1999complex}, and sociology \cite{watts1998collective}. It has been established that systems of coupled entities can exhibit various synchronization patterns \cite{abrams2016introduction}. For example, recent studies of adaptive networks have revealed the phenomenon of recurrent synchronization \cite{thiele2023asymmetric}, a macroscopic event characterized by a periodic transition between synchronous and asynchronous behaviors. The underlying mechanisms of recurrent synchronization are attributed to the recurrent slow dynamics of hidden variables related to the coupling weights, especially the asymmetry of adaptation rules. 
A more general chaotic recurrent clustering has been reported in \citep{salesRecurrentChaoticClustering2024}.
The concept of cluster synchronization \cite{belykh2008cluster,liu2016finite,della2020symmetries} where the network is divided into synchronous sub-populations or clusters, is able to provide a framework for the understanding of dynamical coherence in coupled oscillatory systems. These clusters can dynamically form, dissolve, and even exhibit complex behaviors, such as synchronization cluster bursting, reflecting the rich interactions within the network. Bursting occurs when a system trajectory undergoes alternations between a rest state (silent phase) and an active state (repetitive spiking) \cite{rinzel2006bursting,izhikevich2000neural,han2011bursting}. In synchronization cluster bursting, different synchronization states appear and disappear in a burst-like manner, which has been frequently observed in various non-adaptive models \cite{batista2010delayed,fan2017synchronization,li2023transitions}. However, synchronization cluster bursting in adaptive networks has not yet been thoroughly investigated.

In this paper, we study the dynamics of an adaptive oscillator network based on an adaptive phase oscillator model. We conduct an analysis of the general model of $N$ oscillators to reveal its key properties and symmetries. In particular, a new order parameter different from the standard Kuramoto order parameter is proposed which accounts for the symmetries of the model and allows us to better characterize synchronization. Then, we focus on the numerical investigation of the model of three adaptive phase oscillators, and demonstrate the distinct transitions between cluster synchronization and global synchronization. A useful quantity in this context is the time-dependent average frequency difference, which is used to identify the cluster synchronization state as a function of time. We discover that a quiescent state of the order parameter corresponds to global frequency synchronization, while oscillatory behavior indicates partial frequency synchronization. The transition between these two states gives rise to the effect of  synchronization cluster bursting. Furthermore, we simplified the adaptive system into a reduced normal form equation. By conducting a systematic bifurcation and stability analysis of the normal form equation, we uncover the underlying mechanisms that give rise to synchronization cluster bursting.

\section{Model of adaptively coupled oscillators}

In this section, we will give a brief introduction to the adaptive phase oscillator model. First, we describe the general adaptive phase oscillator model with $N$ oscillators in subsection \ref{sec:General model}. Then, we study symmetries of the general model in subsection \ref{sec:Symmetries of the general model}. Finally, in subsection \ref{sec:Order parameter}, we introduce an order parameter that respects the symmetries of the model.

\subsection{General adaptive phase oscillator model}
\label{sec:General model}
We begin this study by introducing a paradigmatic system of $N$ coupled phase oscillators with adaptive coupling \cite{seliger2002plasticity,Aoki2009,kasatkin2017self}, which is given by
\begin{equation}\label{eq: general model}
\begin{array}{l}
\dot{\varphi}_{i}=\omega_{i}-\frac{1}{N}\sum_{j=1}^{N}\kappa_{ij}\sin(\varphi_{i}-\varphi_{j}), \\
\dot{\kappa}_{ij}=-\varepsilon[\kappa_{ij}+A_{ij}\sin(\varphi_{i}-\varphi_{j}+\delta_{ij})].
\end{array}
\end{equation}
Here ${\varphi _i} \in \mathbb{T}^1$ denotes the phase and ${\omega _i}$ is the natural frequency of oscillator $i$ ($i=1,..,N$).  We use the notation $\mathbb{T}^N$ for the $N$ dimensional torus.  The interactions between the oscillators are quantified through the adaptive coupling weights, ${\kappa _{ij}}$, which represent the connection strength from the $j$th to the $i$th oscillator, where $i,j=1,\dots,N$, $i\ne j$. The positive parameter $\varepsilon$ is the rate of adaptation of the coupling weights. The parameters ${\delta _{ij}}$ are phase-lags of the adaptation function, and the coupling parameters $A_{ij}>0$ represent an underlying topology.  Therefore, we have an $N^2$--dimensional phase space $
(\varphi_1,\dots,\varphi_N, \kappa_{1,2},\dots,\kappa_{1,N},\dots,\kappa_{N,1},\dots,\kappa_{N,N-1})\in
\mathbb{T}^N\times\mathbb{R}^{N(N-1)} $.

We note that in the following we will choose the parameter $\varepsilon $ to be small, to obtain the case where the  adaptation of the coupling weights is slow compared to the rapid dynamics of the phase oscillators. Moreover, the coupling weights remain confined within intervals $ - A_{ij} \le {\kappa _{ij}} \le A_{ij}$ due to the existence of the attracting region 
$G = \left\{ {\left( {{\varphi _i},{\kappa _{ij}}} \right)\in \mathbb{T}^N\times\mathbb{R}^{N(N-1)} :\ \left|\kappa _{ij} \right| \le A_{ij},\ i,j = 1, \ldots ,N} \right\}$ \cite{kasatkin2017self}.
In the following, we will study the model in the case of $A_{ij}\leq1$ and $0<\epsilon <1$.

\subsection{Symmetries of the general model}
\label{sec:Symmetries of the general model}
The model introduced in the previous section allows for the following non-trivial phase-space symmetries, which are crucial for the analysis of the dynamics and synchronization.
\subsubsection{Continuous phase shift symmetry}\label{sec:Phase shift symmetry of the general model}
We note from the model equations (\ref{eq: general model}) that only phase differences appear on the right side of the model equations. Therefore, the form of the equations will remain unchanged, if we add a constant phase shift to all oscillators. That is, the model has a continuous phase shift symmetry
\begin{equation}\label{eq: Phase-Shift-General model}
\gamma _\sigma ^{\, c} : \ {\varphi _i} \mapsto {\varphi _i} + \sigma
\end{equation}
for all $i = 1, \ldots ,N$ and $\sigma  \in {\mathbb{T}^1}$.

\subsubsection{Discrete phase space symmetry}
From equations (\ref{eq: general model}) we observe that a phase shift by $\pi$ in the $l$th oscillator will change the signs of all sine terms which contain $\varphi_l$. This sign change is compensated, if we then also change the signs of all those $\kappa_{ij}$ with $i=l$ or $j=l$. 
More precisely, the system possesses the discrete phase space symmetries $\gamma_l$ ($l=1,\dots,N$), given by
\begin{equation} \label{eq: discrete phase space symmetry}
\begin{aligned}
\gamma _l ^{\, d}: \ &\left(\varphi_l;\ \kappa_{1l},\ldots,\kappa_{Nl},\kappa_{l1},\ldots,\kappa_{lN}\right)\mapsto\\&\left(\varphi_l+\pi;\ -\kappa_{1l},\ldots,-\kappa_{Nl},-\kappa_{l1},\ldots,-\kappa_{lN}\right).
\end{aligned} 
\end{equation}
Note that in (\ref{eq: discrete phase space symmetry}) we only indicate the $2N-1$ elements which are changed, and all other $(N-1)^2$ variables remain unchanged.
It is clear that $\gamma _l^{\, d} \circ \gamma _l^{\, d} =e$, where $e$ is the identical operation, and $\gamma_l^{\, d}\circ\gamma_m^{\, d} = \gamma_m^{\, d}\circ\gamma_l^{\, d}$ for $l \ne m$. Therefore, the symmetry group generated by the $\gamma_l$ has $2^{N}$ elements and is isomorphic to $(\mathbb{Z}_2)^N=\mathbb{Z}_2\times ... \times \mathbb{Z}_2$. The continuous and discrete symmetries are connected through the relation $\gamma _\pi ^{\, c} = \gamma _1^{\, d} \circ \gamma _2^{\, d} \circ  \cdots  \circ \gamma _N^{\, d}$.

\subsection{Order parameter}
\label{sec:Order parameter}
The standard Kuramoto order parameter $\tilde{R}(t)$ is given by \cite{kuramoto1984chemical}
\begin{equation}\label{eq: standard OR}
    \tilde R(t)= \frac{1}{N}\left| {\sum\limits_{j = 1}^N {{e^{{\rm{i}}{\varphi _j}(t)}}} } \right|,
\end{equation}
which is invariant under the symmetry operation $\gamma _\sigma^{\, c}$. However, it is not invariant under the discrete phase space symmetry operation $\gamma _l^{\, d}$, since in this case one of the terms in equation (\ref{eq: standard OR}) changes sign. We therefore propose a different order parameter $R(t)$ which respects both discrete and continuous phase space symmetries, given by
\begin{equation}
R(t) = \frac{1}{{N(N - 1)}}\left| {\sum\limits_{\substack{i,j = 1\\
 i \ne j}}^N {{\kappa _{ij}}{e^{i({\varphi _i} - {\varphi _j})}}} } \right|.
\end{equation}
This order parameter obeys the two symmetry relations $R(t) \circ \gamma _\sigma ^{\, c} = R(t)$ and $R(t) \circ \gamma _l^{\, d} = R(t)$ as desired and also fulfills the constraint
\begin{equation}
    0 \leq R(t) \leq 1.
\end{equation}
$R(t)$ can be interpreted as an averaged "synaptic input" of the system, since it is an average of the real parts of the coupling terms in the phase dynamics from \eqref{eq: general model}.

\section{Three adaptive phase oscillators}
\label{sec:Three oscillator model}
\begin{figure}
    \centering
    \includegraphics[width=0.5\columnwidth]{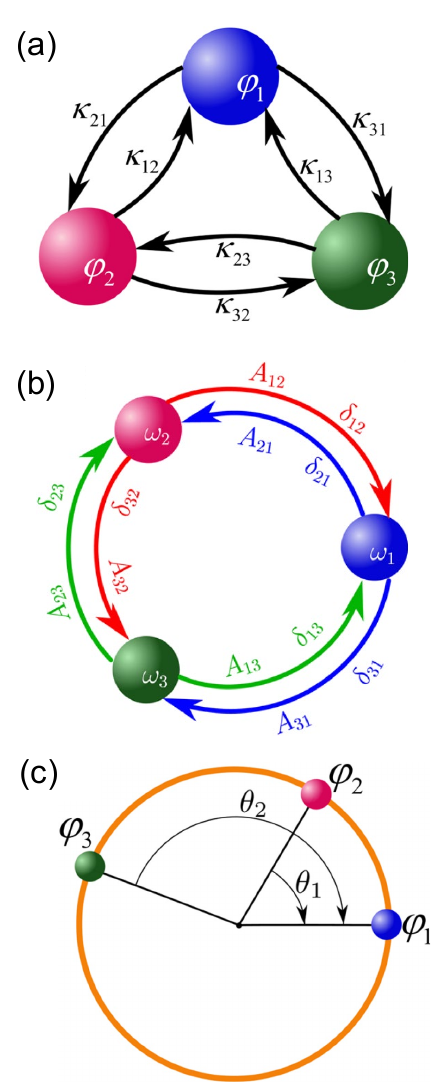}
    \caption{(a) Schematic representation of the three coupled phase oscillators. (b) The oscillator network. (c) The phase differences.}
    \label{fig: Three Osc Scheme}
\end{figure}

To investigate synchronization cluster bursting in adaptive oscillators networks, we first focus on the case of three coupled phase oscillators. As shown in Fig.~\ref{fig: Three Osc Scheme} (a), the three oscillators $\varphi_1$, $\varphi_2$ and $\varphi_3$ are connected by the six adaptive coupling weights $\kappa_{ij}$ ($i,j=1,2,3$, $i\ne j$). The coupling parameters of the network are illustrated in Fig.~\ref{fig: Three Osc Scheme} (b).

\subsection{System in phase difference variables}
\label{sec:System in phase differences}

It is convenient to reduce the dimensionality of the model by introducing phase difference variables as illustrated in Fig.~\ref{fig: Three Osc Scheme} (c).  We thus define 
\begin{equation}\label{eq: Phase_Differences}
    \theta_1 = \varphi_1 - \varphi_2, 
    \qquad
    \theta_2 = \varphi_1 - \varphi_3,
\end{equation}
and  the frequency differences
\begin{equation}\label{eq: Frequency_Differences}
    \Delta_1 = \omega_1 - \omega_2, 
    \qquad
    \Delta_2 = \omega_1 - \omega_3.
\end{equation}
The original three phase oscillators model is transformed into the phase difference model
\begin{equation}
\label{eq: Reduced Syst}
\begin{split}
{{\dot \theta }_1} = &\Delta_1 - [({\kappa _{12}} + {\kappa _{21}})\sin {\theta _1} + {\kappa _{13}}\sin {\theta _2} \\
&+ {\kappa _{23}}\sin ({\theta _1} - {\theta _2})]/3\\
{{\dot \theta }_2} = &\Delta_2- [{\kappa _{12}}\sin {\theta _1} + ({\kappa _{13}} + {\kappa _{31}})\sin {\theta _2} \\
&+ {\kappa _{32}}\sin ({\theta _2} - {\theta _1})]/3\\
{{\dot \kappa }_{12}} =  &- \varepsilon [{\kappa _{12}} + {A_{12}}\sin ({\theta _1} + {\delta _{12}})]\\
{{\dot \kappa }_{13}} =  &- \varepsilon [{\kappa _{13}} + {A_{13}}\sin ({\theta _2} + {\delta _{13}})]\\
{{\dot \kappa }_{21}} =  &- \varepsilon [{\kappa _{21}} + {A_{21}}\sin ( - {\theta _1} + {\delta _{21}})]\\
{{\dot \kappa }_{23}} =  &- \varepsilon [{\kappa _{23}} + {A_{23}}\sin ({\theta _2} - {\theta _1} + {\delta _{23}})]\\
{{\dot \kappa }_{31}} =  &- \varepsilon [{\kappa _{31}} + {A_{31}}\sin ( - {\theta _2} + {\delta _{31}})]\\
{{\dot \kappa }_{32}} =  &- \varepsilon [{\kappa _{32}} + {A_{32}}\sin ({\theta _1} - {\theta _2} + {\delta _{32}})].
\end{split}
\end{equation}

\subsection{Klein group symmetry \texorpdfstring{$\mathbf{K}_4$}{K4}} 
\label{Klein group symmetry}
While the symmetry operations $\gamma^c_\sigma$ translates into the identity operation in the phase difference model,  the original symmetries $\gamma_l^d$ translate into non-trivial symmetries $\tilde \gamma _1^d$  as follows:

\begin{equation}
\begin{split}
   \tilde \gamma _1^{\, d} : & \quad (\theta_{1},\,\theta_{2},\,\kappa_{12},\,\kappa_{13},\,
   \kappa_{21},\,\kappa_{23},\,\kappa_{31},\,\kappa_{32})\,\longmapsto
   \\
&(\theta_{1}+\pi,\,\theta_{2}+\pi,\,-\kappa_{12},\,-\kappa_{13},\,-\kappa_{21},\,\kappa_{23},\,-\kappa_{31},\,\kappa_{32}),
\end{split}
\end{equation}
\begin{equation}
\begin{split}
   \tilde \gamma _2^{\, d} : & \quad (\theta_{1},\,\theta_{2},\,\kappa_{12},\,\kappa_{13},\,   \kappa_{21},\,\kappa_{23},\,\kappa_{31},\,\kappa_{32})\,\longmapsto
\\
&(\theta_{1}+\pi,\,\theta_{2},\,-\kappa_{12},\,\kappa_{13},\,-\kappa_{21},\,-\kappa_{23},\,\kappa_{31},\,-\kappa_{32}),
\end{split}
\end{equation}
\begin{equation}
\begin{split}
   \tilde \gamma _3^{\, d} : & \quad (\theta_{1},\,\theta_{2},\,\kappa_{12},\,\kappa_{13},\,  \kappa_{21},\,\kappa_{23},\,\kappa_{31},\,\kappa_{32})\,\longmapsto
   \\
&(\theta_{1},\,\theta_{2}+\pi,\,\kappa_{12},\,-\kappa_{13},\,\kappa_{21},\,-\kappa_{23},\,-\kappa_{31},\,-\kappa_{32}).
\end{split}
\end{equation}

These three actions together with the identity element $e$ generate the Klein group of four elements \cite{armstrong1988groups}: $\{e,\tilde \gamma _1^{\, d}, \tilde \gamma _2^{\, d},\tilde \gamma _3^{\, d}\}$, which is commutative and has the properties $\tilde \gamma _i^{\, d} \circ \tilde \gamma _i^{\, d}  = e$ and $\tilde \gamma _1^{\, d} \circ \tilde \gamma _2^{\, d} \circ \tilde \gamma _3^{\, d} = e$. The Klein-four group can be also represented as the direct product: $K_4=\mathbb{Z}_2\times\mathbb{Z}_2$.

\section{Numerical observation of synchronization cluster bursting}
\label{sec:Numerical observation of SCB}
In this section, we numerically demonstrate the existence of synchronization cluster bursting for the case of three adaptively coupled phase oscillators. We introduce the average frequency difference as a measure to determine the synchronization state among different oscillators of the system. Finally, we show how the symmetry of the system allows us to infer the existence of a second synchronization cluster state.

\subsection{Synchronization cluster bursting}\label{sec:Synchronization cluster bursting} 

Here, we focus on the three oscillators adaptive phase model (\ref{eq: Reduced Syst}). In Fig.~\ref{fig:R_t_pi} and \ref{fig:theta_t_pi} we show a typical numerical scenario for parameters given in Appendix \ref{Initial values and parameter values}.  It is observed that the system exhibits periodic changes from fast repetitive spiking to an episode of steady dynamics, which can be regarded as bursting oscillations. At a given time $t$, the level of synchronization is measured by the order parameter $R(t)$. As shown in Fig.~\ref{fig:R_t_pi} (a), the order parameter transitions between two distinct states: a nearly constant value, representing the rest state of bursting, and a strongly oscillating order parameter indicating the active state of bursting.  From Fig.~\ref{fig:R_t_pi} (b) we see that the various coupling weights act quite differently during the transition between the oscillating and the rest state.  While $\kappa_{23}$ and $\kappa_{32}$ remain approximately constant throughout, the other $\kappa_{ij}$ engage in a low frequency oscillatory behaviour around zero.  When $R(t)$ oscillates, the coupling weights $\kappa_{ij}$ also exhibit rapid fluctuations, but at a very small amplitude.  In comparison, the oscillations of the phase differences $\theta_{1,2}$ presented in Fig.~\ref{fig:theta_t_pi} are large during the oscillation phase of the order parameter. This indicates the presence of distinct time scales, with the coupling weights being the slow and the phase differences being the fast variables. 

The observed changes  of the order parameter over time in Fig.~\ref{fig:R_t_pi} (a) as well as the evolution of the phase differences in Fig.~\ref{fig:theta_t_pi} indicate transitions in the system's synchronization states.  We would therefore characterize the time-dependent synchronization state of the system. Let us define the average frequency $\bar \Omega_i(t )$ of oscillator $i$ at time $t$ over some interval from $t-\Delta t/2$ to $t+\Delta t/2$  as
\begin{equation}
{{\bar \Omega }_i}(t ) 
 = \frac{{{\varphi _i}(t-\frac{\Delta t}{2}) - {\varphi _i}(t+\frac{\Delta t}{2} )}}{\Delta t},
\label{eq:phasediff3d}
\end{equation}
where we choose a suitable $\Delta t$ which is smaller than the timescale of the adaptive weights, but larger than the typical timescale of the phase variables.  For our case, we choose  $1 \le \Delta t \le \frac{1}{\varepsilon }$.
The frequency differences are then calculated as follows:
\begin{equation}
{\bar \Omega }_{ij}(t) = \left|{\bar \Omega }_i(t) - {\bar \Omega }_j(t)\right|.
\end{equation}
We now stipulate that oscillators $i$ and $j$ are frequency synchronized at time $t$ if the frequency difference is below some threshold parameter $\eta$, i.e. ${\bar \Omega }_{ij}(t) < \eta$. In this work we chose $\eta =1.0\times 10^{-3}$.  We then say that a set $\mathcal{C} = \{i_1,\ldots,i_k\}$ forms a synchronization cluster, if each pair of oscillators in the set are frequency synchronized, i.e. if $\bar{ \Omega}_{ij}(t) < \eta$ for all $i,j \in \mathcal{C}$. In the case of three oscillators, the notation  $\{ 1,2,3\} $ therefore indicates global frequency synchronization \cite{belykh2008cluster}, and the notation $\{ i,j\} ,k$ indicates partial frequency synchronization between oscillators $i$ and $j$, but oscillator $k$ not being part of the synchronization cluster.

Using the frequency difference ${{\bar \Omega }_{ij}(t)}$, we segment $R(t)$ into distinct sections, as shown in Fig.~\ref{fig:R_t_pi}(a). Here, the blue line represents global frequency synchronization denoted by $\{1,\ 2,\ 3\}$, while the red line indicates partial frequency synchronization denoted by $\{2,\ 3\},\ 1$. $R(t)$ periodically undergoes transitions between global synchronization and cluster synchronization  in a burst-like manner. Such phenomena can be classified as synchronization cluster bursting. In particular, before global frequency synchronization terminates and partial frequency synchronization begins, $R(t)$ shows several distinct small oscillations. Moreover, it is observed from Figs.~\ref{fig:theta_t_pi} (a) and (b) that during global synchronization, the phase differences remain nearly constant. Conversely, in states of partial synchronization, the phase difference rotates between 0 and $2 \pi$. Similarly, small oscillations also occur when the synchronization switches from global synchronization to partial synchronization. Finally, we would like to point out that the phase difference between $\varphi_2$ and $\varphi_3$ is almost constant, i.e., ${\theta _2} - {\theta _1} \approx \pi $  (see Fig.~\ref{fig:theta_t_pi} (c)).

\subsection{Alternative synchronization cluster bursting due to symmetry}\label{sec:Deriving alternative SCB}

\begin{figure}
    \centering
    \includegraphics[width=1.00\columnwidth]{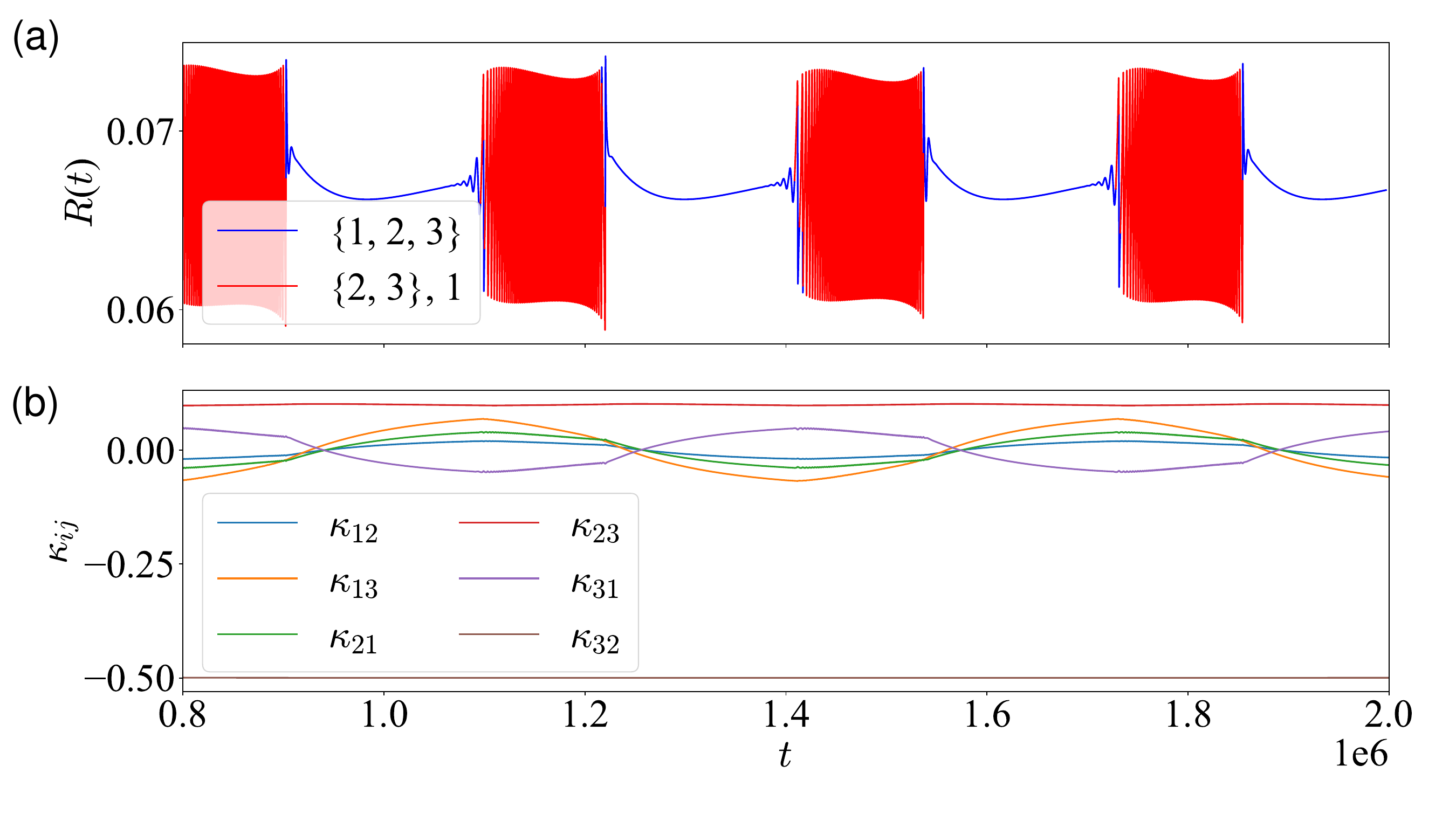}\\
    \includegraphics[width=0.48\columnwidth]{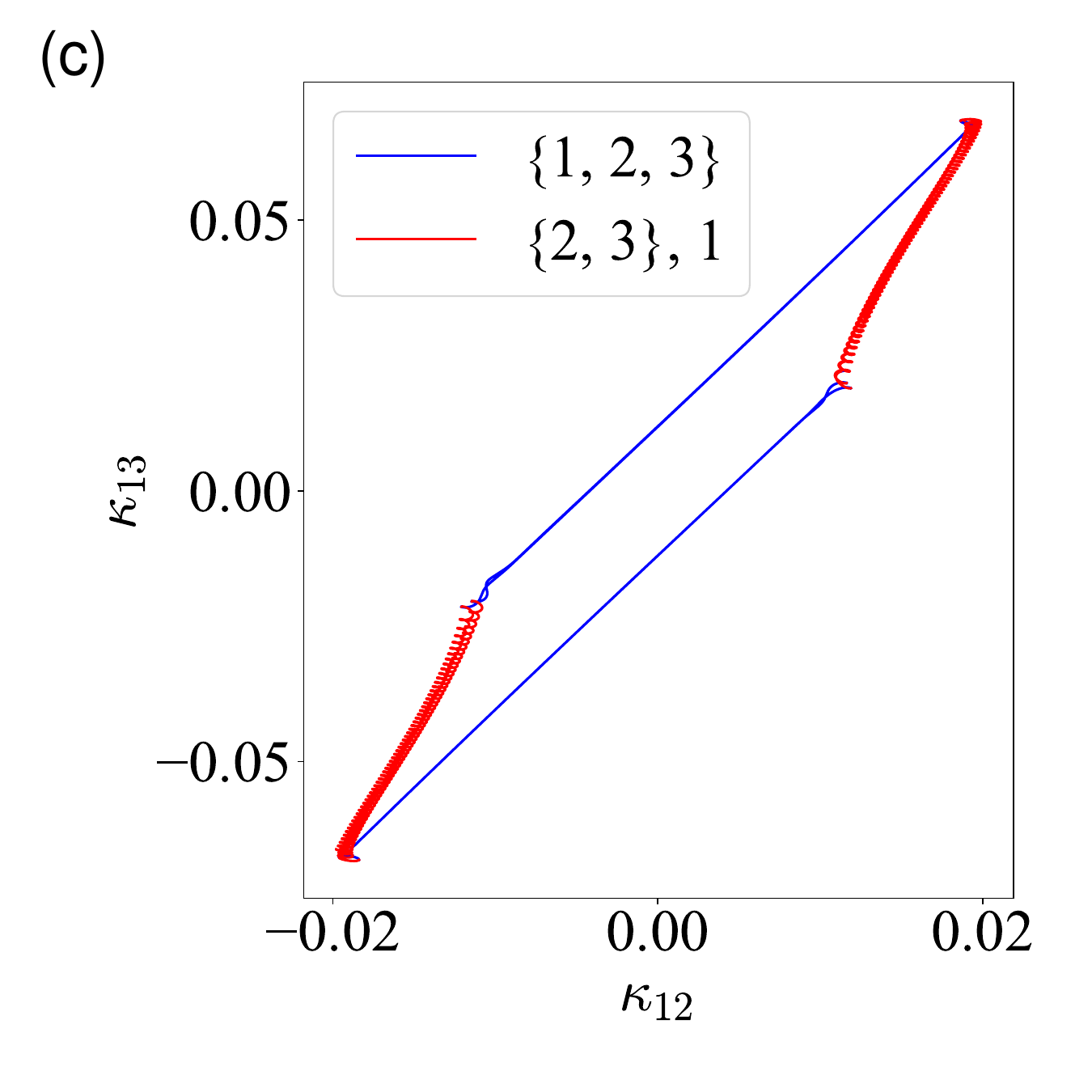}
    \includegraphics[width=0.48\columnwidth]{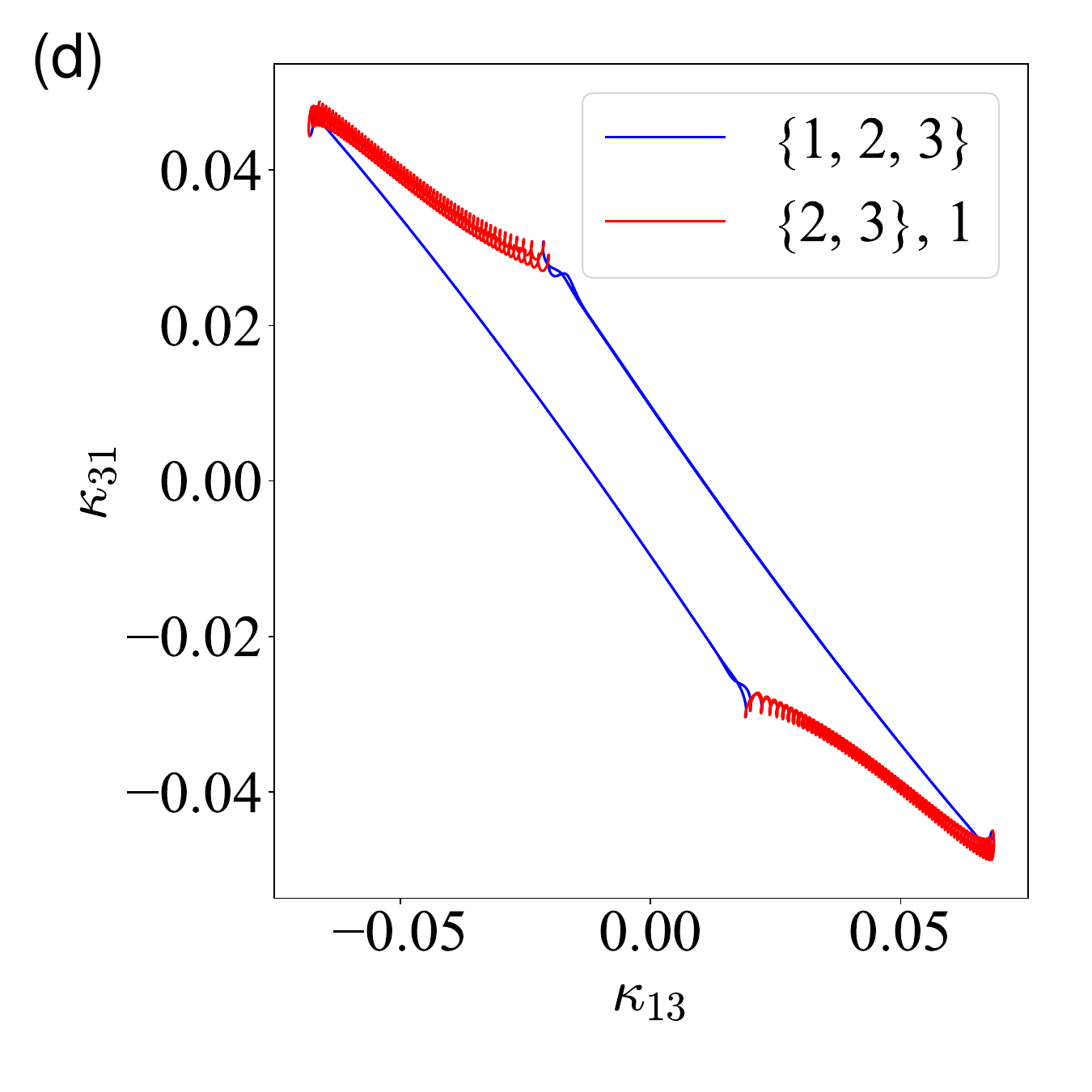}
    \caption{Synchronization cluster bursting in three adaptive phase oscillators model (\ref{eq: Reduced Syst}) Time series of order parameter $R$ (a) and coupling weights $\kappa_{ij}$ (b). Phase portraits in the coupling variables ($\kappa_{12}$, $\kappa_{13}$) (c) and ($\kappa_{13}$, $\kappa_{31}$) (d). The blue line, i.e., $\{1,2,3\}$, represents the three oscillators that are globally synchronized, while the red line, i.e., $\{2,3\},1$, means oscillators 2 and 3 are synchronized. All parameters are provided in Appendix \ref{Initial values and parameter values}. The initial values of the variables can be selected randomly. To replicate this figure, the initial values can be found in Appendix \ref{Initial values and parameter values}.} 
    \label{fig:R_t_pi}
\end{figure}

\begin{figure}
    \centering
    \includegraphics[width=1.00\columnwidth]{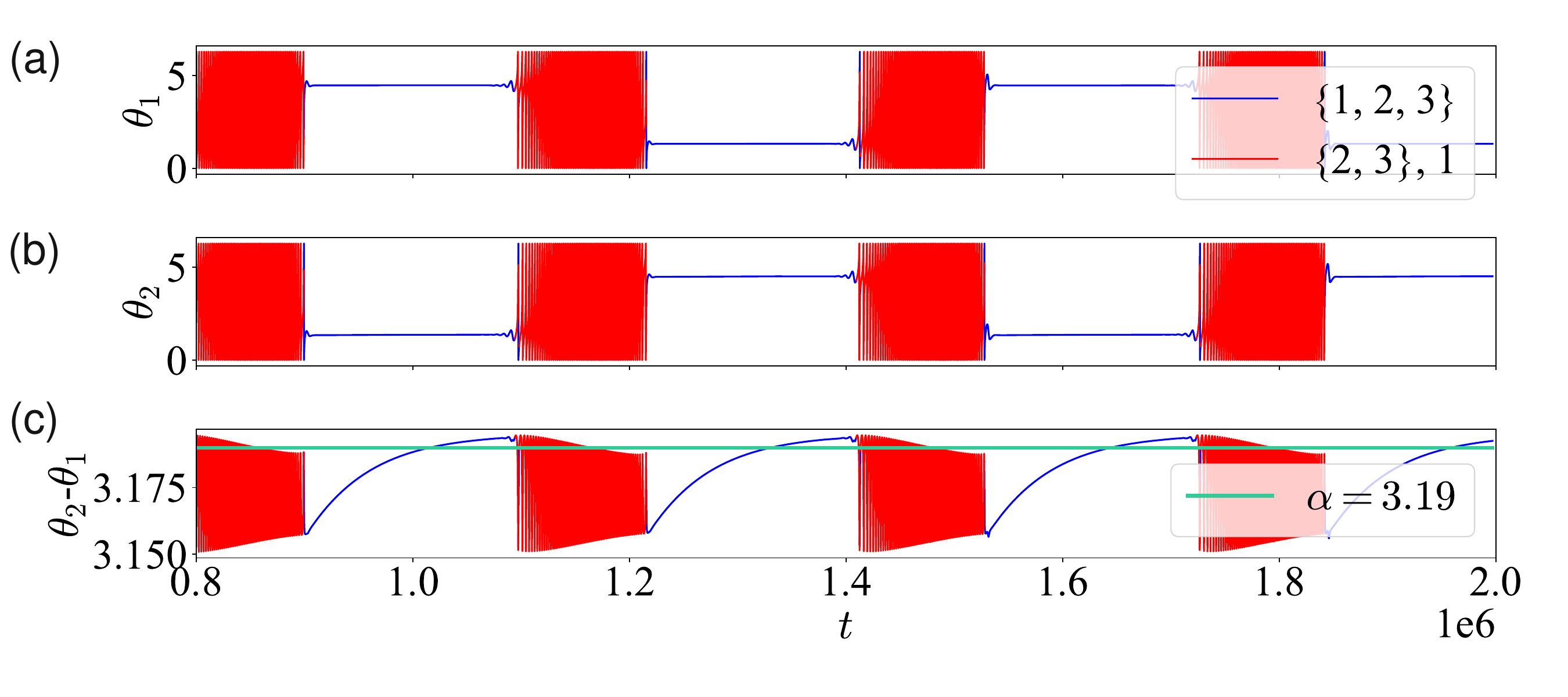}
    \caption{Dynamics of the system in phase differences for three oscillators (\ref{eq: Reduced Syst}). Time series of $\theta_{1}$ (a), $\theta_{2}$ (b), and $\ \theta_{2} - \theta_{1}$ (c).} 
    \label{fig:theta_t_pi}
\end{figure}

\begin{figure}
    \centering
    \includegraphics[width=1.00\columnwidth]{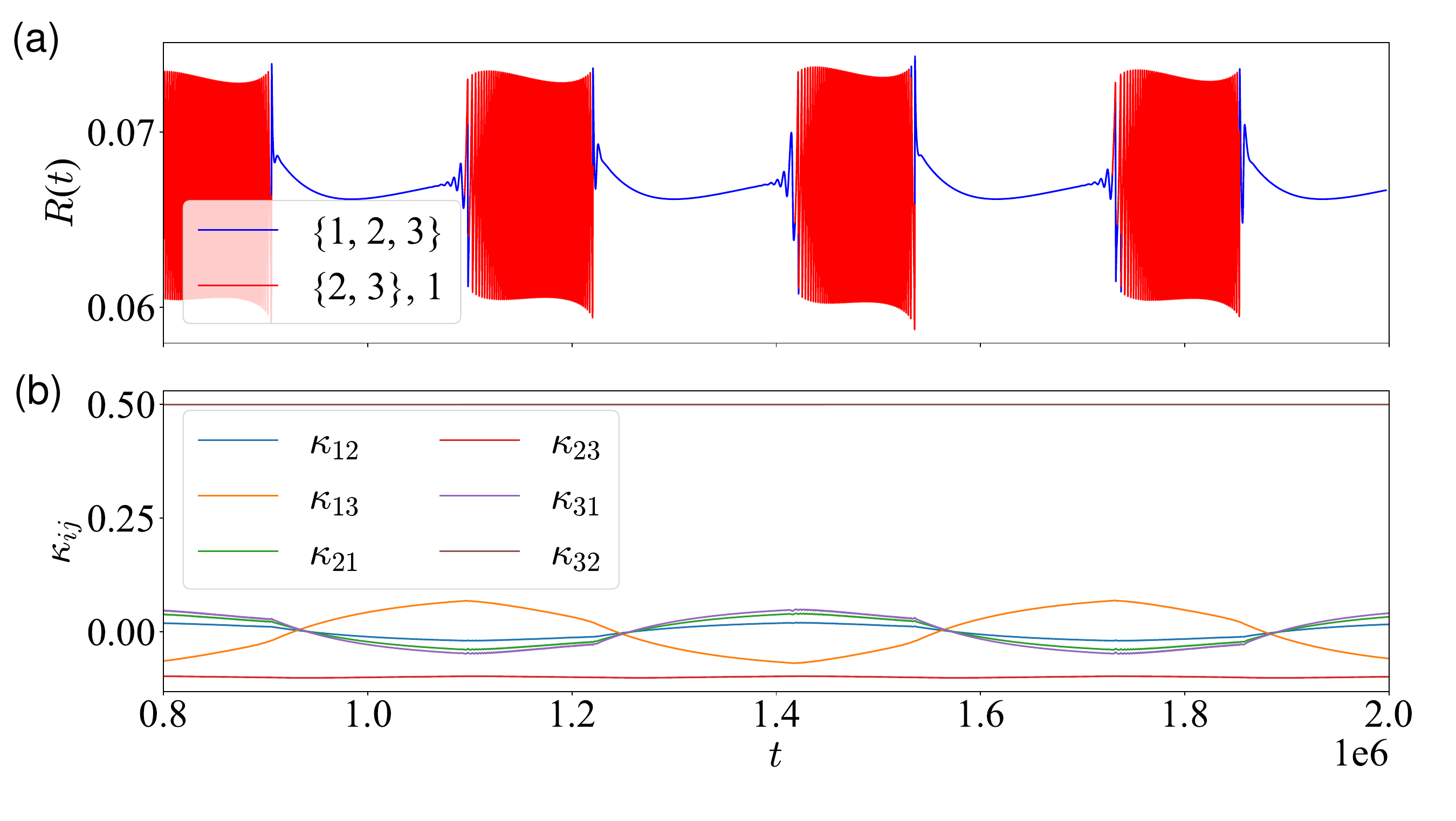}\\
    \caption{The same as Fig.~\ref{fig:R_t_pi}. Another synchronization cluster bursting pattern obtained through symmetry action $\tilde{\gamma}_{2}^d$.}
    \label{fig:R_t_0}
\end{figure}

\begin{figure}
    \centering
    \includegraphics[width=1.00\columnwidth]{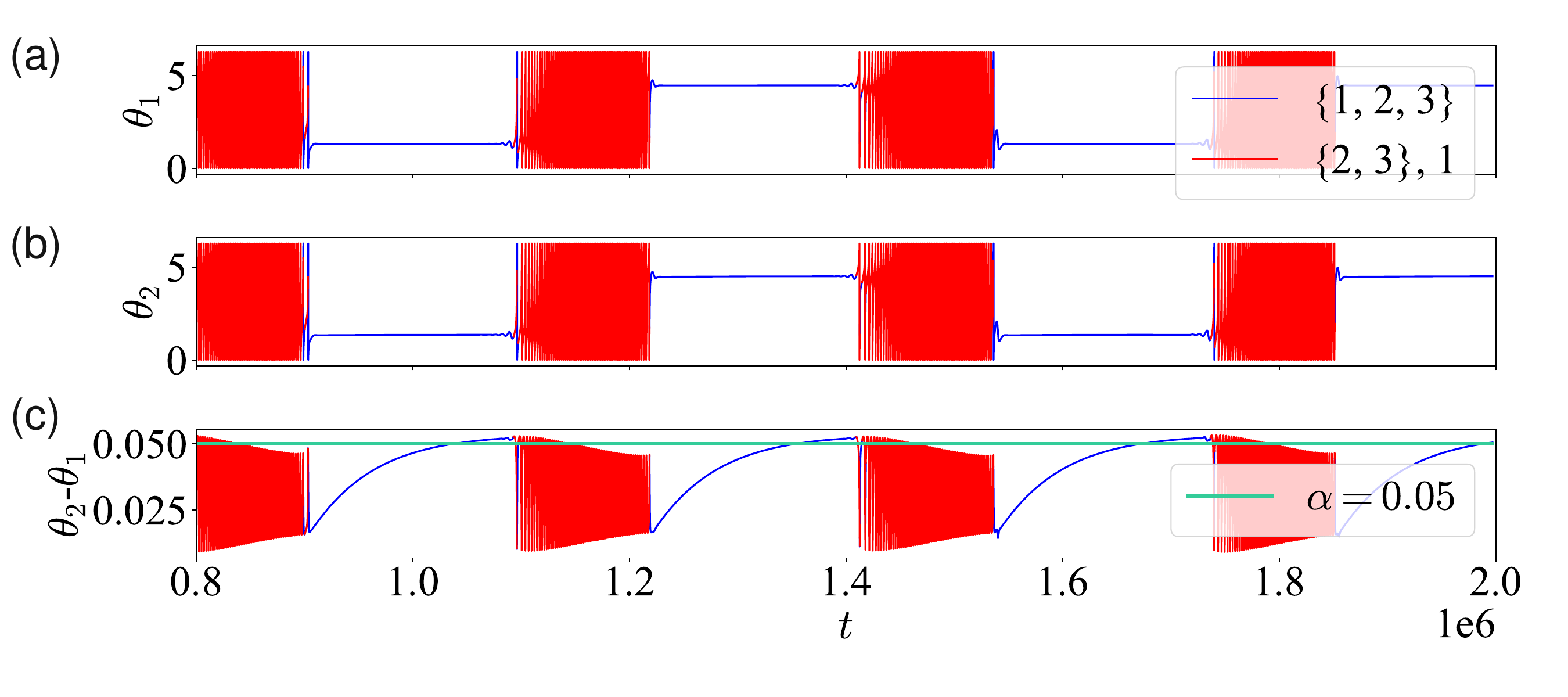}
    \caption{The same as Fig.~\ref{fig:theta_t_pi} for the case in Fig.~\ref{fig:R_t_0}.} 
    \label{fig:theta_t_0}
\end{figure}

The dynamics shown in Figs.~\ref{fig:R_t_pi} and \ref{fig:theta_t_pi} is not the only stable attractor of the system, and indeed with different initial conditions we find a second stable attractor as shown in Figs.~\ref{fig:R_t_0} and \ref{fig:theta_t_0}. A comparison of the dynamics in either case reveals that applying the symmetry operation $\tilde{\gamma}_{2}^d$ provides a way to transition between the two attractors. This symmetry operation in particular flips the signs of the almost constant $\kappa_{23}$ and $\kappa_{32}$, and thereby create a manifestly different dynamical state. It also changes $\theta_1$ by $\pi$ and leaves $\theta_2$ invariant, which explain why  ${\theta _2} - {\theta _1} \approx 0 $ in Fig.~\ref{fig:theta_t_0} (c). 

It is interesting to note that the symmetry operation $\tilde{\gamma}_{1}^d$, which changes both $\theta_1$ and $\theta_2$ by $\pi$, does not yield a new attractor when applied to the dynamics of Figs.~\ref{fig:R_t_pi} and \ref{fig:theta_t_pi}. Instead, it induces a time shift by half a period on the same attractor. Similarly, because of $\tilde{\gamma}_2^d\circ\tilde{\gamma}_1^d=\tilde{\gamma}_3^d$, the application of $\tilde{\gamma}_3^d$ to the attractor in Figs.~\ref{fig:R_t_pi} and \ref{fig:theta_t_pi} yields a time-shifted version of the attractor in Figs.~\ref{fig:R_t_0} and \ref{fig:theta_t_0}.

\section{Analytical description of synchronization cluster bursting}
\label{sec:Analytical description of SCB}

In the following, we explain the mechanism of the synchronization cluster bursting observed in the three oscillators adaptive phase model. By reducing the original model to an almost-invariant manifold, in subsection \ref{sec:Reduction to the almost-invariant manifold}, we derive a simplified three-dimensional model, which is transformed into a normal form in subsection \ref{sec:Normal form equation}. The investigation of fixed points and bifurcations of the normal form equation enhances our understanding of the origin of the cluster bursting phenomena.

\subsection{Reduction to an almost-invariant manifold}
\label{sec:Reduction to the almost-invariant manifold}

\begin{figure}
    \centering
    \includegraphics[width=0.8\columnwidth]{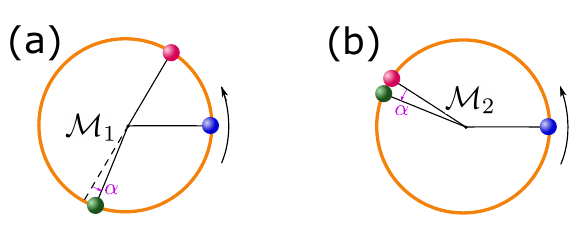}
    \caption{Schematic representation of the phase difference between $\varphi_2$ and $\varphi_3$}
    \label{fig:Phases difference alpha}
\end{figure}

\begin{figure}
    \centering
    \includegraphics[width=1.00\columnwidth]{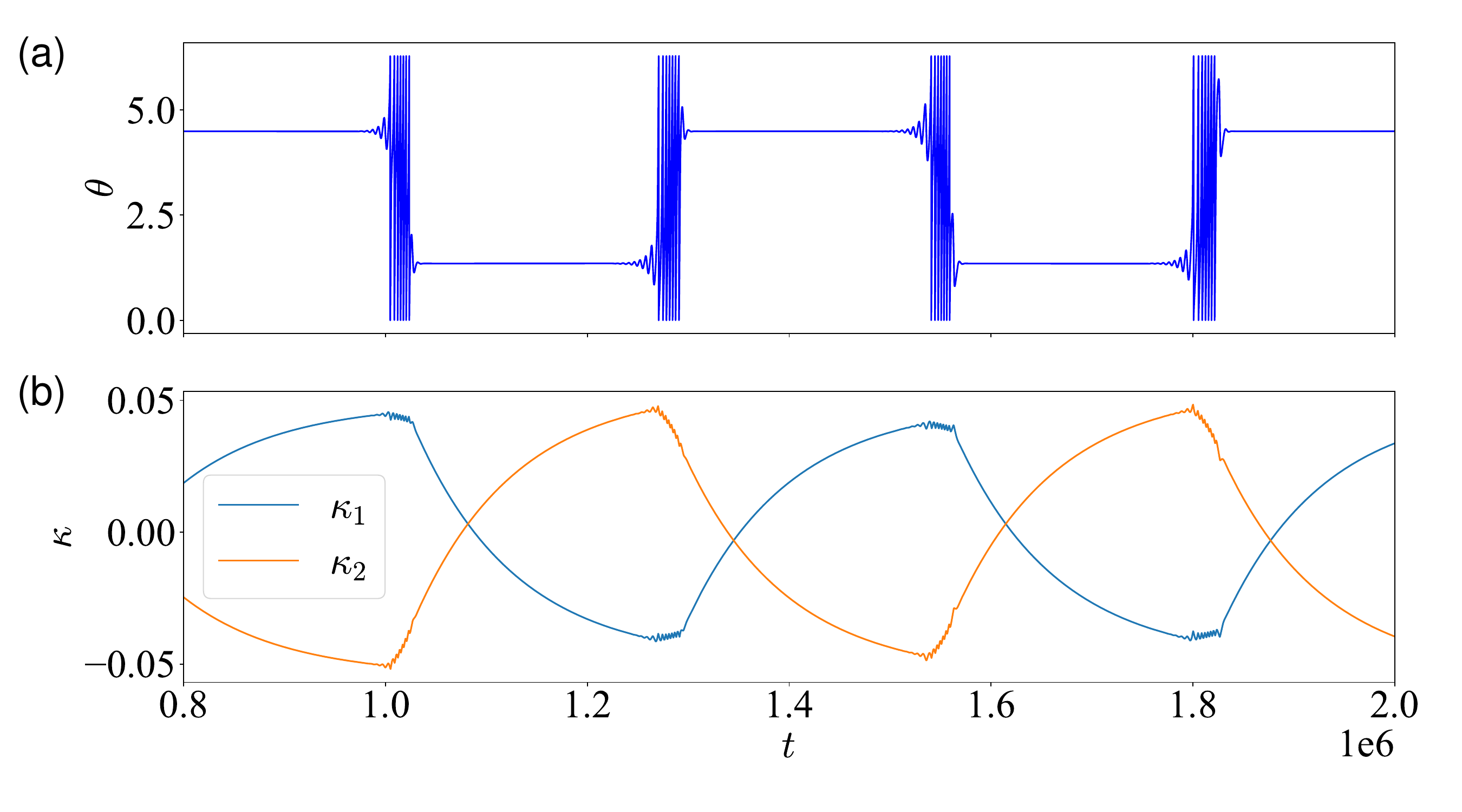}\\
    \includegraphics[width=0.7\columnwidth]{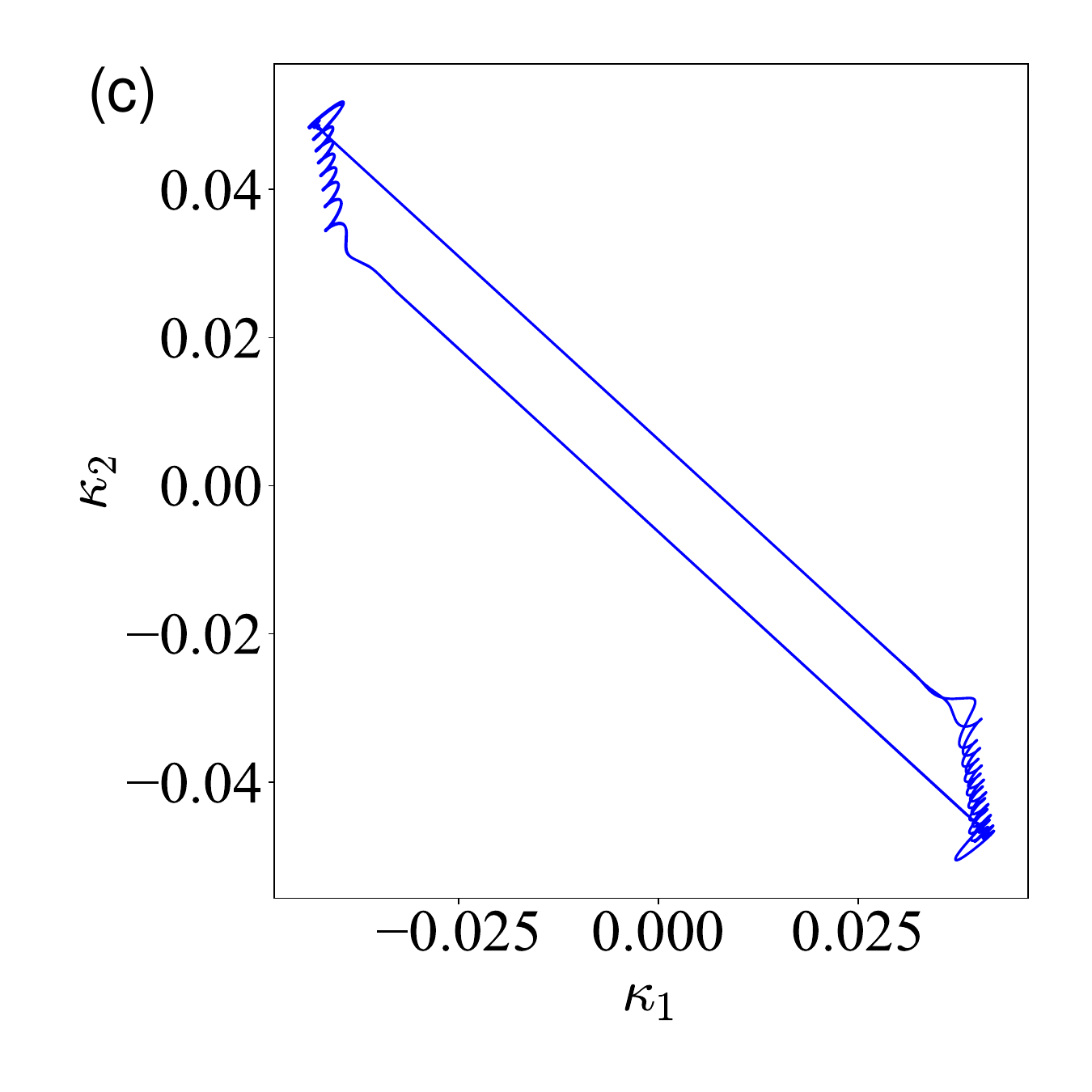}
    \caption{Dynamics of the reduced three-dimensional system (\ref{eq: simple 3-D system}) for the case when $\alpha$ is fixed at 0.05 corresponding to the dynamics in Fig.~\ref{fig:R_t_0}.} 
    \label{fig:3D}
\end{figure}

Based on the numerical observation, that in the cluster bursting attractors the phase difference $\phi_2 - \phi_3 = \theta_2 - \theta_1$ is either close to $\pi$ (see Fig.~\ref{fig:theta_t_pi} (c)) or close to $0$ (Fig.~\ref{fig:theta_t_0} (c)) let us use the approximation
\begin{equation}
\label{eq: phase difference between phi2 and phi3}
{\varphi _2}(t) - {\varphi _3}(t) = \alpha,
\end{equation}
where $\alpha$ is a positive parameter closed to $\pi$ or 0 (see Fig.~\ref{fig:Phases difference alpha}). 
Based on this, we can simplify the original three oscillators model by replacing the phase variables ${\varphi _1}$, ${\varphi _2}$ and ${\varphi _3}$ with ${\varphi _1}$ and ${\frac{{{\varphi _2} + {\varphi _3}}}{2}}$. Then, by introducing the phase difference variable $\theta$ and the frequency difference parameter $\Delta$ via 
\begin{equation}
\label{eq: phase difference in 3D system}
{\varphi _1} - {\frac{{{\varphi _2} + {\varphi _3}}}{2}} = \theta,
    \qquad
{w_1} - ({w_2} + {w_3})/2 = \Delta,
\end{equation}
we obtain the reduced three-dimensional system
\begin{equation}
\label{eq: simple 3-D system}
\begin{array}{l}
\dot \theta  = \Delta  - [{\kappa _1}\sin (\theta  - \frac{\alpha }{2}) + {\kappa _2}\sin (\theta  + \frac{\alpha }{2}) + \beta \sin (\alpha )]/3\\
{{\dot \kappa }_1} =  - \varepsilon [{\kappa _1} + {A_1}\sin (\theta  - \frac{\alpha }{2} + {\delta _1})]\\
{{\dot \kappa }_2} =  - \varepsilon [{\kappa _2} + {A_2}\sin (\theta  + \frac{\alpha }{2} + {\delta _2})].
\end{array}
\end{equation}
The expressions for the parameters $A_{1,2}$, $\delta_{1,2}$ and $\beta$ in this reduced three-dimensional system (\ref{eq: simple 3-D system}) and details can be found in Appendix \ref{APPENDIX A}. As shown in the numerical result in Fig.~\ref{fig:3D}, the dynamics of the reduced model retains the interesting bursting features of the original model, thus validating the effectiveness of the proposed reduction method.

\subsection{Normal form equation}
\label{sec:Normal form equation}
In order to facilitate the subsequent analysis of the underlying mechanisms of synchronization cluster bursting, we transform the model (\ref{eq: simple 3-D system}) to a normal form with a reduced number of parameters. 

Note that 
\begin{equation}
\begin{split}
    &\kappa_{1}\sin\left(\theta-\frac{\alpha}{2}\right)+\kappa_{2}\sin\left(\theta+\frac{\alpha}{2}\right)\\
    = &\text{Im}\left\{ \left[\kappa_{1}e^{i\left(-\frac{\alpha}{2}\right)}+\kappa_{2}
    e^{i
    \left(+\frac{\alpha}{2}\right)  
    }\right]e^{i\theta}\right\}. 
\end{split}
\end{equation}
The square bracket in the last equation is a complex number, parameterized by $\kappa_{1}$ and $\kappa_{2}$. We can therefore replace the two equations for $\kappa_{1}$ and $\kappa_{2}$ in (\ref{eq: simple 3-D system}) with a single complex dynamical quantity. Then, by rescaling the equation, we transform system (\ref{eq: simple 3-D system}) into the normal form
\begin{equation}\label{eq: normal form}
\begin{split}
\dot{\theta} & = 1 - \left| Q \right| \sin\left( \theta - \phi_{Q} \right) \\
\dot{Q} & = -\hat{\epsilon} \left[ Q + C_{1}e^{i\theta} + C_{2}e^{-i\theta} \right],
\end{split}
\end{equation}
where $C_{1},C_{2}\in\mathbb{C}$ and  $\hat{\epsilon}\in\mathbb{R}$ are parameters, and $\theta\in {\mathbb{T}^1}$,  $Q\in\mathbb{C}$ are the dynamical variables, $\phi_{Q}$ represents the argument of $Q$ (see Appendix \ref{APPENDIX B} for more details). 

\subsection{Dynamics of the normal form equation}
\label{sec:Dynamics of the normal form}

With the above simplifications and reductions, we obtained a normal form for synchronization cluster bursting. The underlying mechanism for this effect are now revealed by analyzing the stability and bifurcation of the normal form equation (\ref{eq: normal form}) (see Appendix \ref{APPENDIX C} for more details). First, we observe that the model allows for the following phase space and parameter symmetries,
\begin{equation}
  \left( {\theta ,\,Q} \right) \mapsto \left( {\theta  + \pi ,\, - Q} \right),  
\end{equation}
\begin{equation}
    \left( {\theta ,\,Q,\,{C_2}} \right) \mapsto \left( {\theta  + \sigma ,\,Q{e^{i\sigma }},\,{C_2}{e^{2i\sigma }}} \right),
\end{equation}
where $\sigma\in {\mathbb{T}^1}$ is an arbitrary angle. As a result, we can change the phase of $C_{2}$ at the cost of adjusting the angle $\theta$. We can use this symmetry to always choose $C_{2}\in\mathbb{R}_{0}^{+}$. Thus we have only four real parameters, i.e., $\hat{\epsilon}$, $\text{Re}C_{1}$, $\text{Im}C_{1}$, and $C_{2}$.

When the condition
\begin{equation}
\left|\text{Im}C_{1}-1\right|\leq C_{2}
\end{equation}
is satisfied (see also (\ref{eq:fpcond}) in Appendix \ref{APPENDIX C}), there are in general four fixed points, with $\theta$ given by
\begin{align}
\theta_{0,1} & =\frac{\arcsin\left(\frac{\text{Im}C_{1}-1}{C_{2}}\right)}{2}\in\left(-\frac{\pi}{4},\frac{\pi}{4}\right],\label{eq: fixed point theta}\\
\theta_{0,2} & =\frac{\pi}{2}-\theta_{0,1}\in\left[\frac{\pi}{4},3\frac{\pi}{4}\right),\\
\theta_{0,3} & =\theta_{0,1}+\pi\in\left(3\frac{\pi}{4},5\frac{\pi}{4}\right],\\
\theta_{0,4} & =3\frac{\pi}{2}-\theta_{0,1}\in\left[5\frac{\pi}{4},7\frac{\pi}{4}\right).
\end{align}
and the corresponding $Q_{0}$ given by
\begin{align}
Q_{0,1} & =-C_{1}e^{i\theta_{0,1}}-C_{2}e^{-i\theta_{0,1}},\\
Q_{0,2} & =-iC_{1}e^{-i\theta_{0,1}}+iC_{2}e^{i\theta_{0,1}},\\
Q_{0,3} & =-Q_{0,1},\\
Q_{0,4} & =-Q_{0,2}.
\label{eq: fixed point Q}
\end{align}

Based on this, we next present the results of the stability and bifurcation analysis of the fixed points, with a detailed analysis provided in Appendix \ref{APPENDIX C}. The result shows that $\theta_{0,1}$ and $\theta_{0,3}$ are saddle points with two unstable directions and one stable eigen-direction $\lambda_{1}=-\hat{\epsilon}$, while the stability of $\theta_{0,2}$ and $\theta_{0,4}$ depends on the sign of
\begin{equation}
    T  =\text{Re}C_{1}+C_{2}\cos2\theta-\hat{\epsilon}.
\end{equation}
If $T<0$, we have a stable node or focus with three stable directions; if $T>0$, we have two unstable directions. The parameter points, where a saddle-node bifurcations occur is given by
\begin{equation}
C_{2}=\left|\text{Im}C_{1}-1\right|.
\end{equation}
At $T=0$, we have a Hopf bifurcation, which can be expressed as
\begin{equation}
\left|C_{1}-(\hat{\epsilon}+i)\right|=C_{2}.
\end{equation}
This defines a circle in the complex $C_1$ plane with stable fixed points on the inside and only unstable fixed points on the outside, where the center of the circle is at $(\epsilon,1)$ and the radius is $C_2$. Note that only one half of the circle corresponds to an actual Hopf bifurcation, while on the other half the eigenvalues are real and have opposite signs.

\begin{figure}
    \centering
    \includegraphics[width=0.95\columnwidth]{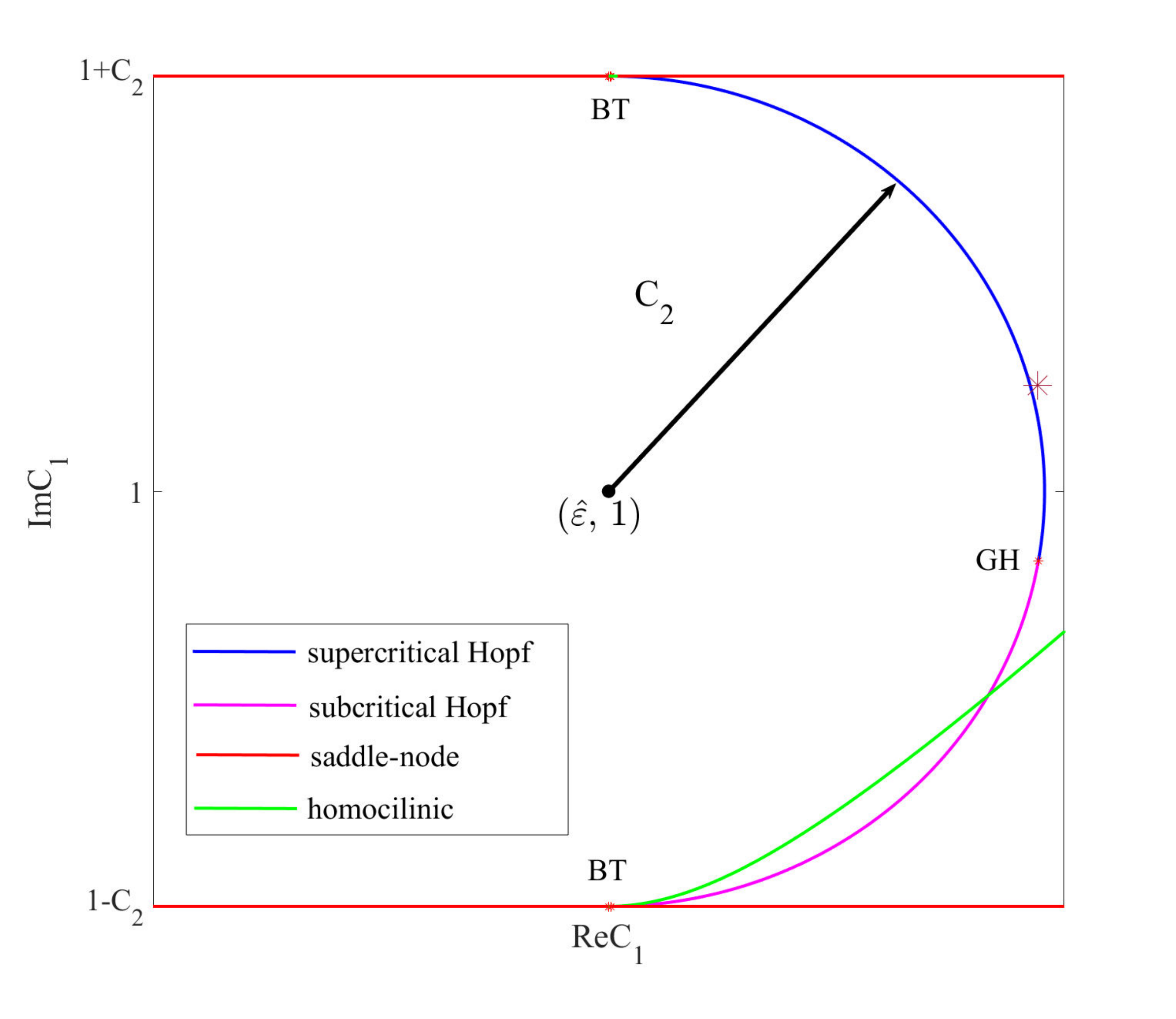}
    \caption{Bifurcation set of the normal form equation (\ref{eq: normal form}) on the parameter plane (${\mathop{\rm Re}\nolimits} {C_1}$, ${\mathop{\rm Im}\nolimits} {C_1}$). The asterisk marks the place where we study.}
    \label{fig:two_parameter_bf}
\end{figure}

Furthermore, we conducted a bifurcation analysis of the normal form equation (\ref{eq: normal form}) using the MatCont software. As shown in Fig.~\ref{fig:two_parameter_bf}, we plot the bifurcation set on the parameter plan (${\mathop{\rm Re}\nolimits} {C_1}$, ${\mathop{\rm Im}\nolimits} {C_1}$), where the Hopf bifurcation curves and homoclinic bifurcation curves terminate at a codimension 2 Bogdanov-Takens (BT) point on the saddle-node bifurcation curves. Additionally,  the supercritical Hopf and subcritical Hopf bifurcations meet at the Generalized Hopf (GH) point. 

 Note that the parameter combination we studied numerically in Section~\ref{sec:Numerical observation of SCB} is located outside of the circle of Hopf bifurcation points (as shown in Fig.~\ref{fig:two_parameter_bf} marked by the asterisk). Thus the system only exhibits unstable fixed points including the two saddle-focus points $\theta_{0,2}$ and $\theta_{0,4}$. These unstable fixed points have one stable eigen-direction $\lambda_{1}=-\hat{\epsilon}$, and the (slowly) stable eigen-direction with which the trajectory approaches the fixed point is given by
\begin{equation}\left.v=\left(\begin{array}{c}0\\\cos\theta\\\sin\theta\end{array}\right.\right).\label{eq:eigen-direction}\end{equation}

\begin{figure}
    \centering
    \includegraphics[width=0.9\columnwidth]{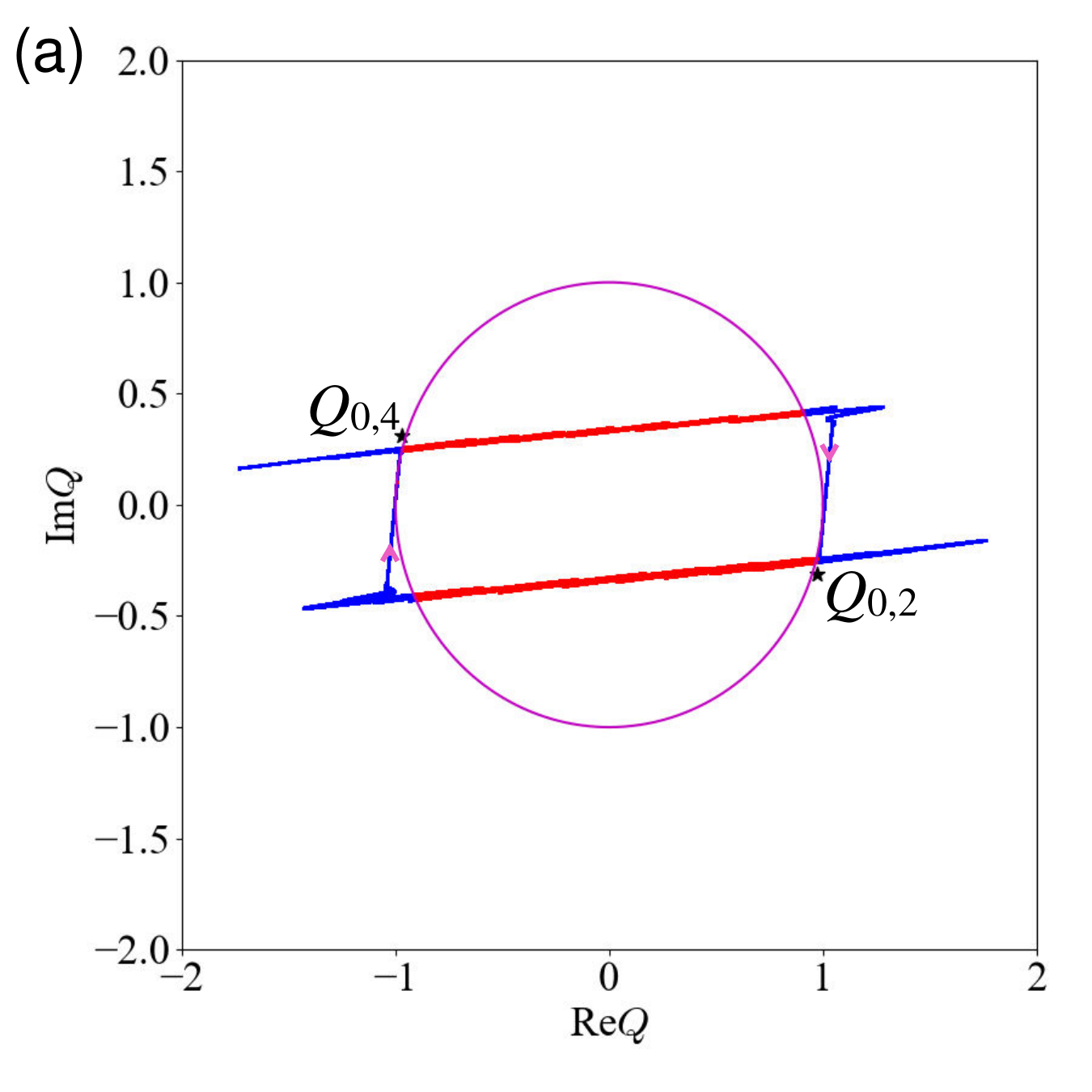}\\
    \includegraphics[width=0.9\columnwidth]{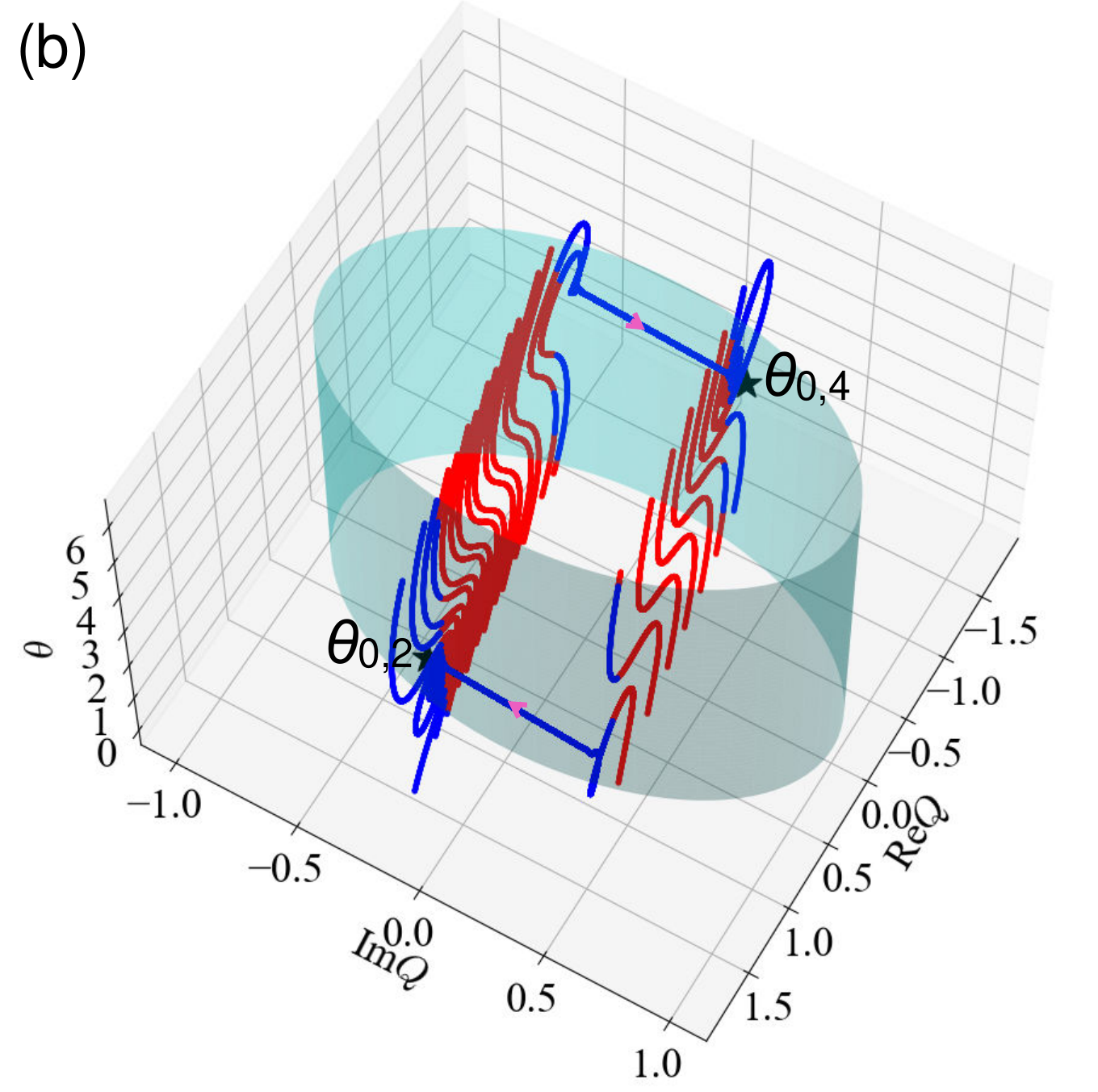}
    \caption{Trajectory of the normal form equation (\ref{eq: normal form}), where $ C_{2}$ is assumed to be a positive real number by taking the modulus of the original complex number $C_{2}$. The blue lines represent trajectories located outside the unit circle or cylinder, while red lines indicate trajectories within it. The pink arrows indicate the direction of motion of the trajectory, which is clockwise. The star markers represent unstable saddle-focus type fixed points.}
    \label{fig:Q}
\end{figure}

\begin{figure}
    \centering
    \includegraphics[width=1.0\columnwidth]{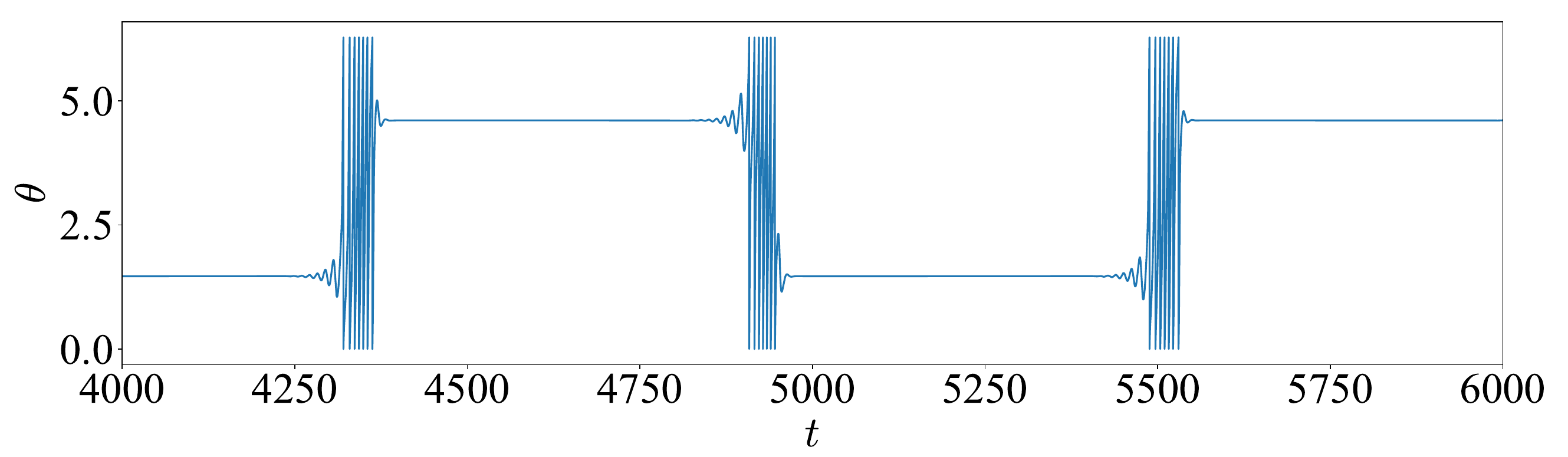}
    \caption{Dynamics of $\theta$ in the normal form equation (\ref{eq: normal form}).}
    \label{fig:theta_t_normal form}
\end{figure}

Figure~\ref{fig:Q} gives an example when $\alpha$ is fixed at $0.05$ corresponding to the synchronization cluster bursting in Figs.~\ref{fig:R_t_0} and \ref{fig:theta_t_0} (when $\alpha =3.19$ corresponding to the synchronization cluster bursting in Figs.~\ref{fig:R_t_pi} and \ref{fig:theta_t_pi}, the situation is also similar). In this case, the parameter values are fixed at ${\mathop{\rm Re}\nolimits} {C_1} =31.64 $, ${\mathop{\rm Im}\nolimits} {C_1} =7.61 $, $C_2=32.12$, and $\hat{\epsilon}=0.005$. It can be seen in Fig.~\ref{fig:Q} (a) that, in the complex $Q$ plane, outside the unit cycle, the trajectory approaches the saddle-node fixed point along the eigen-direction (\ref{eq:eigen-direction}), which is more clearly depicted in the three-dimensional representation shown in Fig.~\ref{fig:Q} (b). This dynamical behavior corresponds to the near-horizontal segments in the time series $\theta$ (see Fig.~\ref{fig:theta_t_normal form}). Because the situation we consider is close to the Hopf bifurcation, the trajectory spirals away from the saddle-focus and exhibits a libration, which explains the small-amplitude oscillations in the time series $\theta$ before lager-amplitude oscillations take place as illustrated in Fig.~\ref{fig:theta_t_normal form}. Then, it engages in a rotational motion ranging from 0 to $2\pi$ inside the cylinder, leading to large oscillations. Subsequently, it is attracted by another saddle-focus point. Based on this, we can explain the dynamics of $\theta$ as shown in Fig.~\ref{fig:theta_t_normal form}, where small oscillatory motions without a full round trip (libration) and large oscillations ranging from 0 to $2\pi$ (rotation) can be observed. As a result, one can conclude that the saddle-focus fixed points $\theta_{0,2}$ and $\theta_{0,4}$ may play a crucial role in the generation of synchronization cluster bursting in Fig.~\ref{fig:R_t_0} (a).

\begin{figure}
    \centering
    \includegraphics[width=0.8\columnwidth]{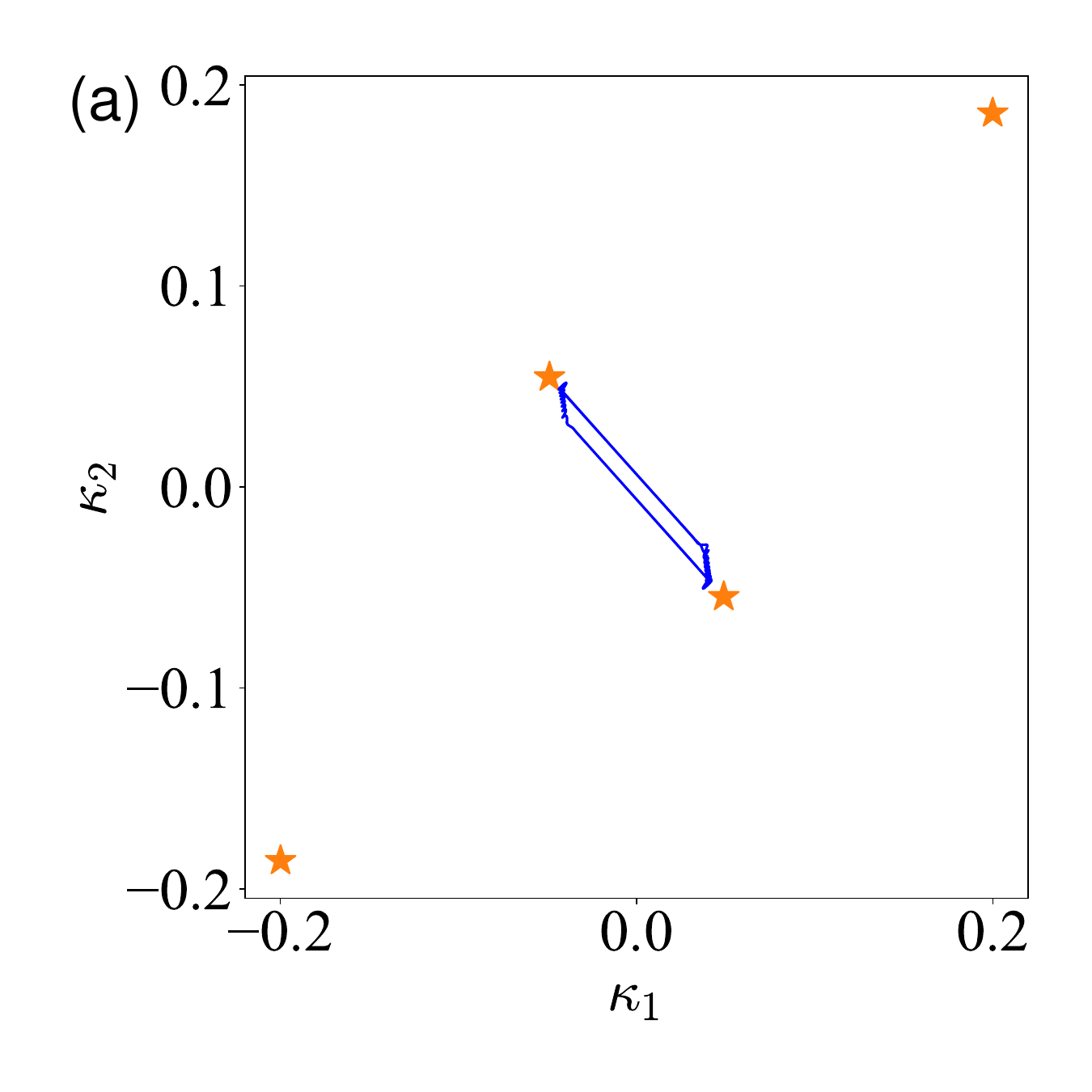}\\
    \includegraphics[width=0.8\columnwidth]{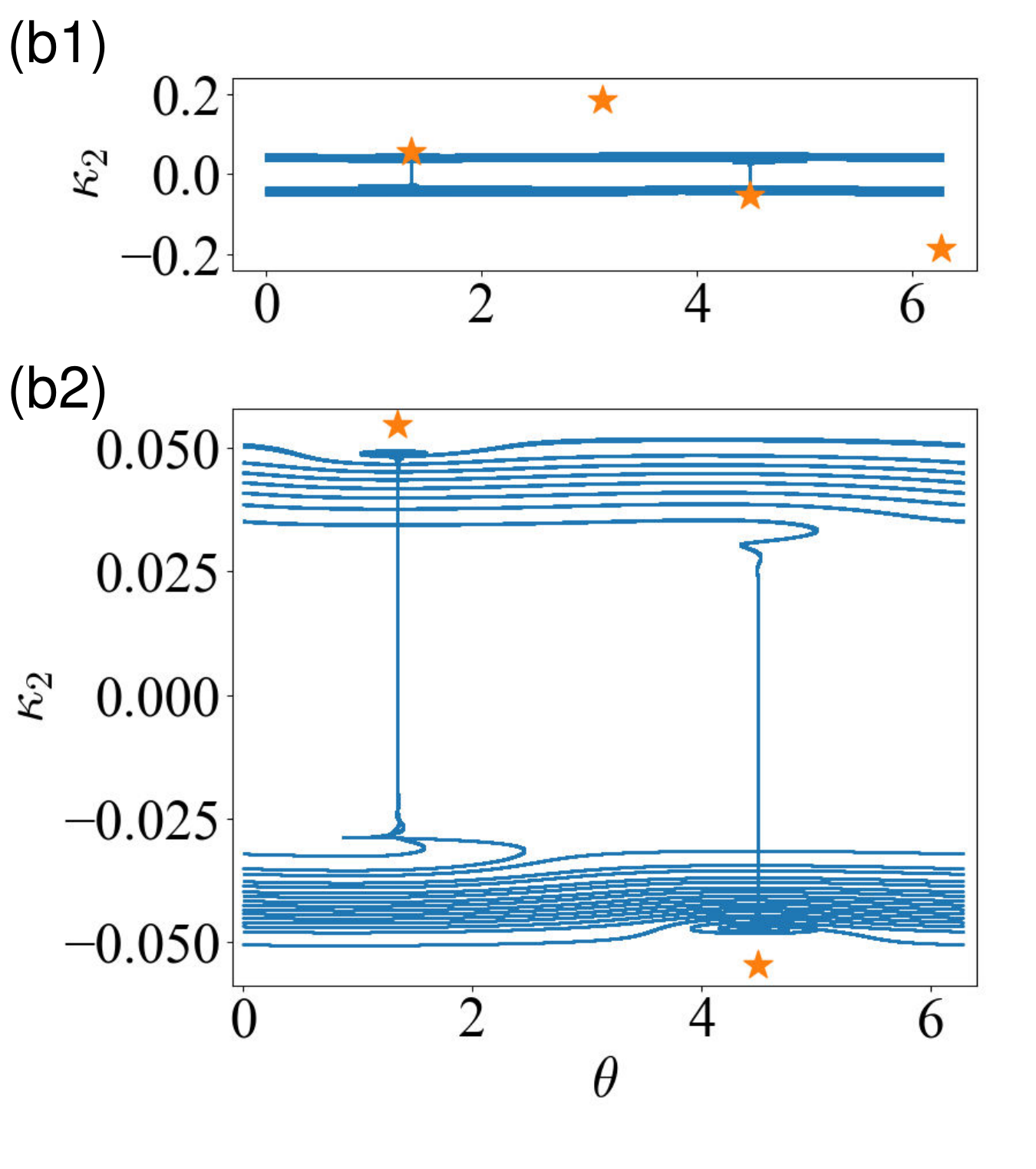}
    \caption{The trajectory of the reduced three-dimensional model (\ref{eq: simple 3-D system}) in the $\kappa_1-\kappa_2$ plane (a), and $\theta-\kappa_2$ plane (b) and its local enlargement (c), in which the fixed points of the reduced three-dimensional model (\ref{eq: simple 3-D system}) is also superimposed.}
    \label{fig:fixed points_3D system}
\end{figure}

Note that the normal form (\ref{eq: normal form}) is derived from the reduced three-dimensional (\ref{eq: simple 3-D system}) through a linear transformation. Consequently, a similar phenomenon is evident in the ($\kappa_1$, $\kappa_2$) plane, where the trajectory also approaches unstable saddle-focus fixed points. Furthermore, this becomes more apparent in the enlarged view of the trajectory on the ($\theta$, $\kappa_2$) plane, as depicted in Fig.~\ref{fig:fixed points_3D system}.

\begin{figure}
    \centering
    \includegraphics[width=0.7\columnwidth]{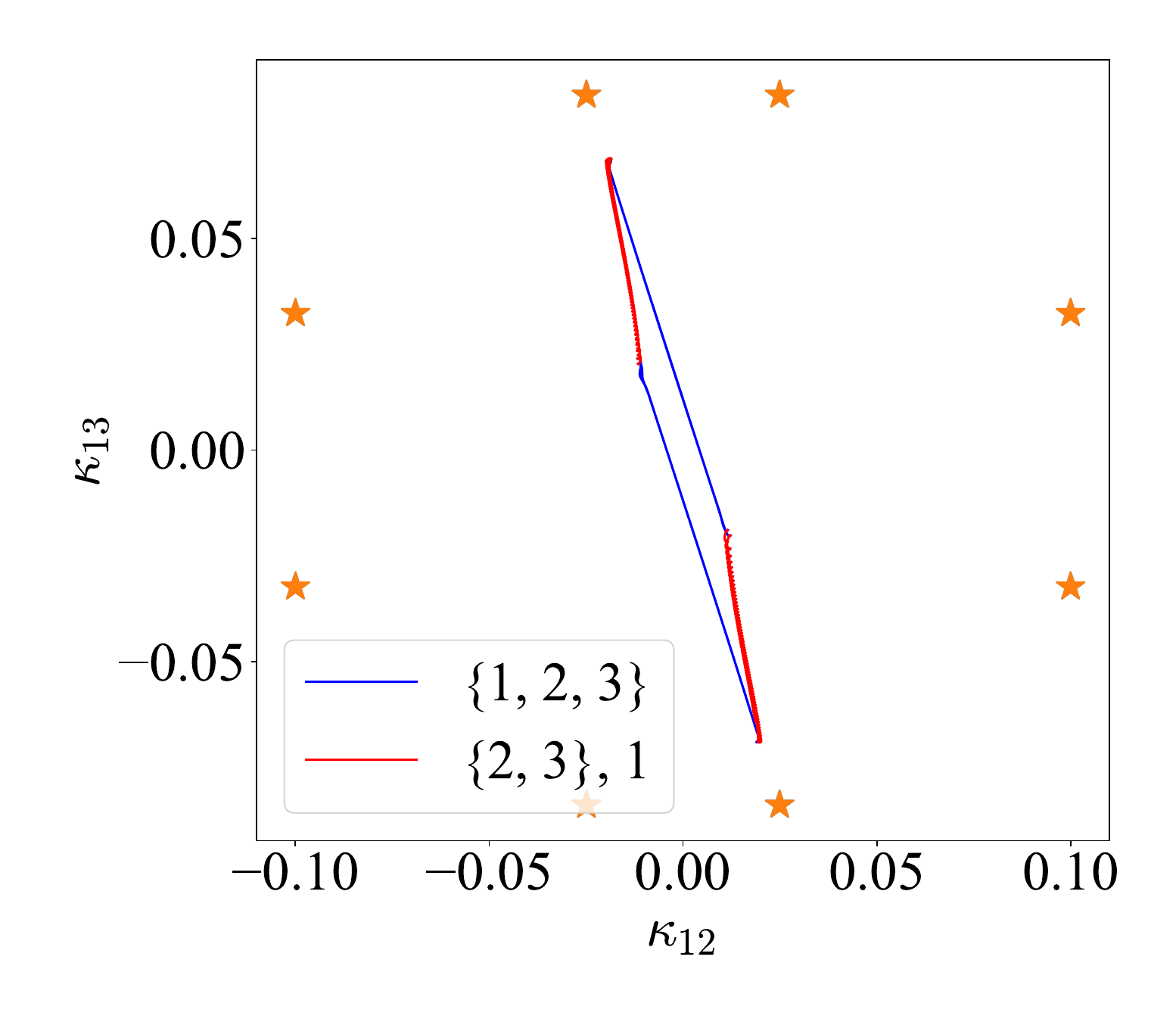}
    \caption{The same as Fig.~\ref{fig:R_t_0} (c), in which the fixed points of the original three oscillators phase model are also superimposed.}
    \label{fig:fixed points_original system}
\end{figure}

Based on the analysis of the normal form equation, we have established that the fixed points, and in particular the saddle points have a significant relevance for its dynamics. Next, the focus will shift to examining the fixed points of the original system (\ref{eq: Reduced Syst}) in the $\kappa$-plane. By numerical fixed points analysis, we find eight fixed points (Fig.~\ref{fig:fixed points_original system}). These are categorized into two groups that can be derived from each other through Klein symmetry, with each group containing four points. Note that in the development of our three-dimensional model, the oscillators ${\varphi _2}$ and ${\varphi _3}$ are assumed to be coupled , which results in a loss of one symmetry, thereby reducing the count to four fixed points. (see Fig.~\ref{fig:fixed points_3D system} (a)). In other words, saddle-node bifurcations occur among the fixed points across the two different groups when the model is reduced.

\section{Conclusion and discussion}
\label{sec:conclusion}
Cluster synchronization has been frequently reported in non-adaptive models. In this paper, we have investigated the cluster synchronization even in adaptive oscillator networks. As a result, the synchronization cluster bursting has been revealed.

We first present a detailed analysis of the general adaptive phase oscillator model, and in particular, we identify the role of continuous phase shift symmetry and discrete phase space symmetry in the dynamics of the system. By introducing a different order parameter respecting both discrete and continuous phase space symmetries, we further quantify the synchronization level of the system.

We focused on the dynamics of the three adaptive phase oscillators and uncovered the influence of the Klein group symmetry $\mathbf{K}_4$ on the system's behavior. Our numerical observations have demonstrated the emergence of synchronization cluster bursting, offering new insights into the mechanisms of synchronization in complex networks. The time-dependent average frequency difference helps us to understand the synchronization cluster bursting, and allows us to observe the system transitions between global and cluster synchronization. Then, we derive alternative synchronization cluster bursting from the Klein group symmetry $\mathbf{K}_4$, which deepens our understanding of synchronization cluster bursting in adaptive oscillators networks.

The reduction to the almost-invariant manifold and the derivation of the normal form equation for the three adaptive phase oscillators provided a more streamlined analytical framework for this complex dynamics. Moreover, we have engaged in a detailed discussion of the dynamics of the normal form to reveal the underlying mechanism of synchronization cluster bursting. The results show that the saddle-focus fixed points play a key role in the emergence of synchronization cluster bursting, near which the system's trajectory undergoes dramatic changes. It first directly approaches the saddle-node fixed point along the eigen-direction. Then, the trajectory spirals away from the saddle-focus creating libration and finally evolving into rotation. Subsequently, the trajectory is drawn to another saddle-focus point. Therefore, the presence of saddle-focus fixed points introduces a mechanism by which the system can transition from a state of global synchronization to one of partial synchronization and back. 

\begin{figure}
    \centering
    \includegraphics[width=0.48\columnwidth]{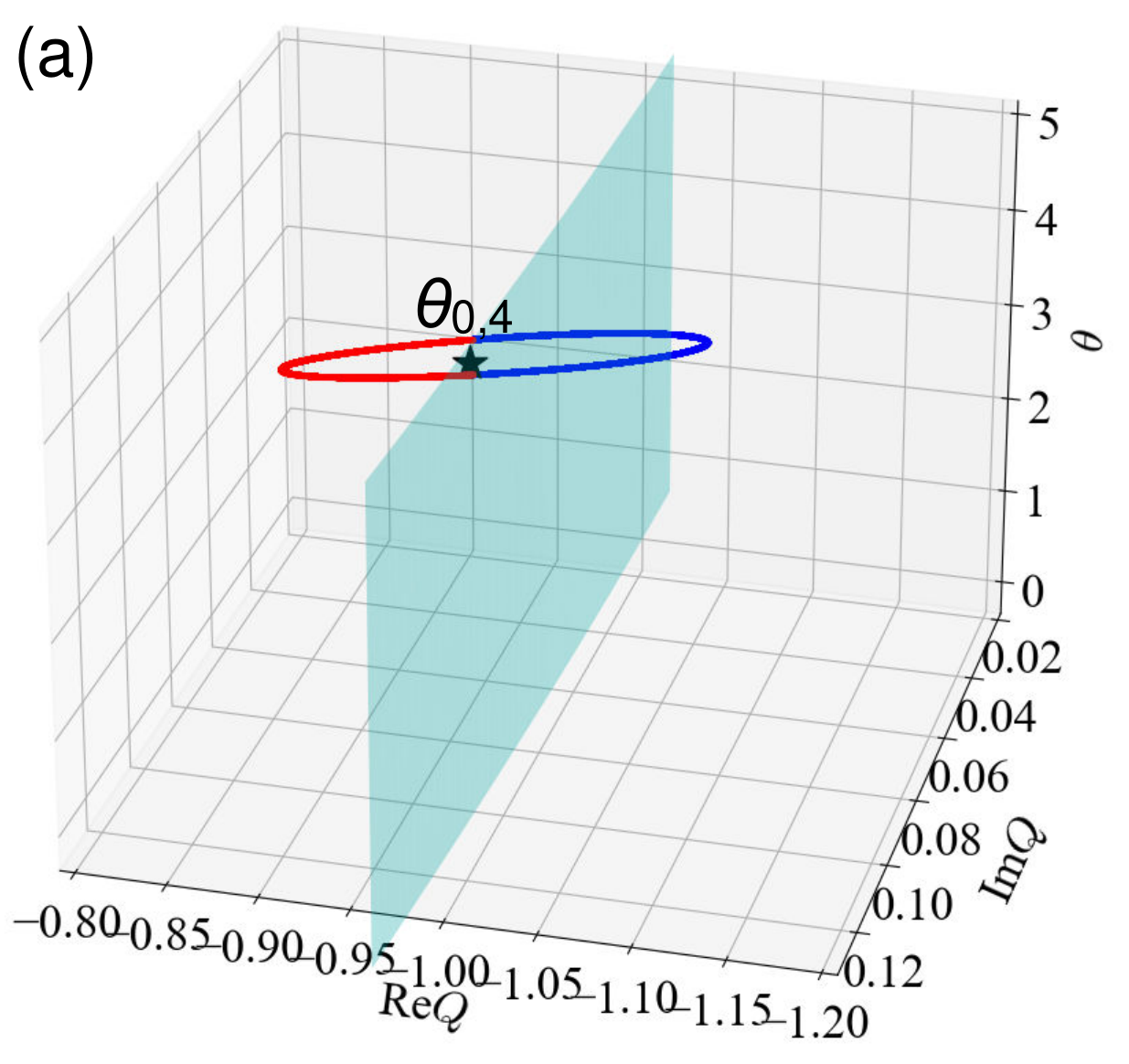}
    \includegraphics[width=0.48\columnwidth]{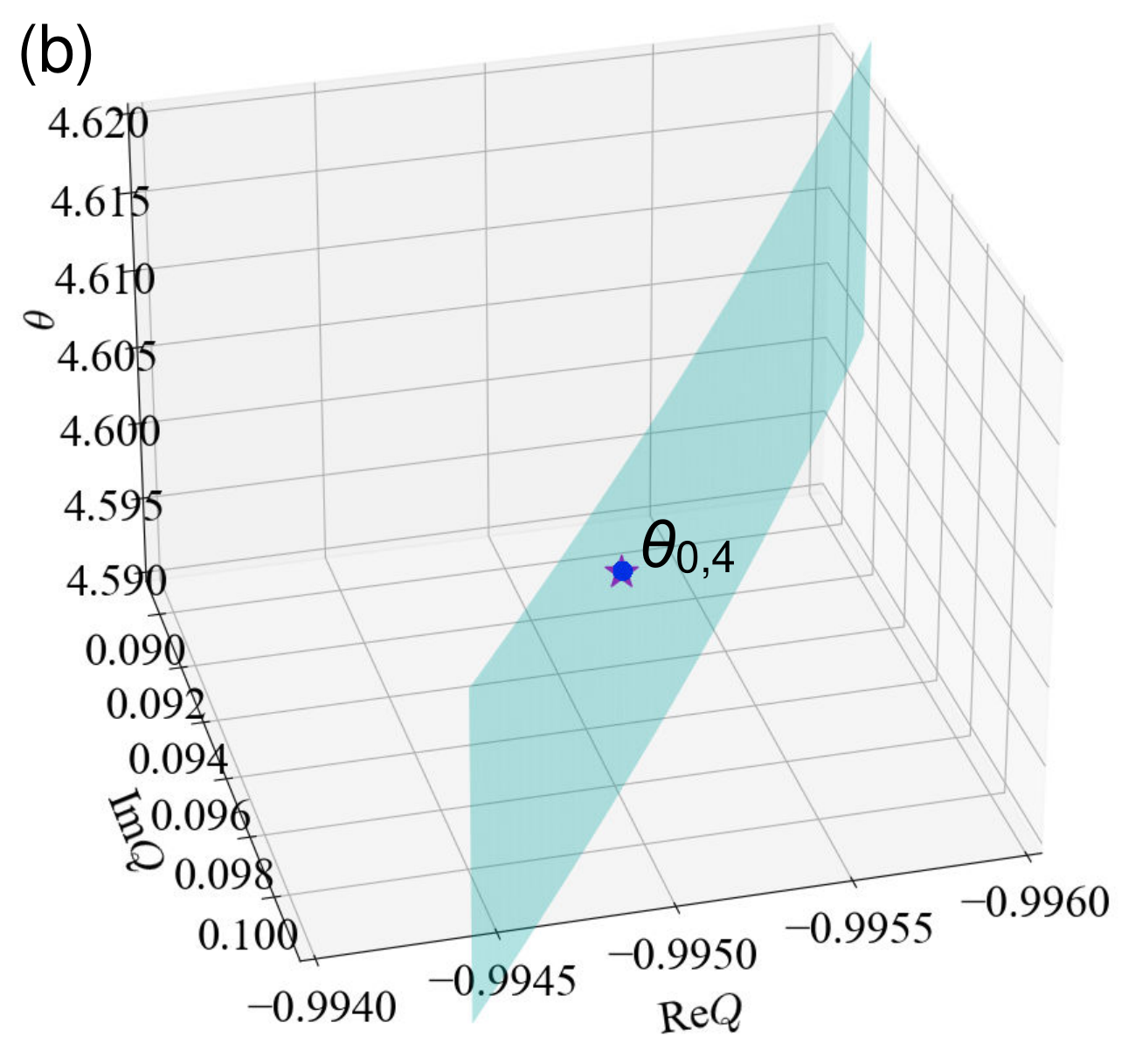}
    \caption{Evolution of the trajectory with the decrease of the ${\mathop{\rm Re}\nolimits} {C_1}$ near the point marked by the asterisk in Fig.~\ref{fig:two_parameter_bf}. (a) ${\mathop{\rm Re}\nolimits} {C_1}=31.44$, $\theta_{0,4}$ is unstable and there exists a stable limit cycle attractor originated from supercritical Hopf bifurcation. (b) ${\mathop{\rm Re}\nolimits} {C_1}=31.43$, $\theta_{0,4}$ is a stable fixed point.}
    \label{fig:Evolution}
\end{figure}

Besides, we would like to point out that variations in parameters induce alterations in the trajectory near the point indicated by the asterisk in Fig.~\ref{fig:two_parameter_bf}. For example, it can be seen in Fig.~\ref{fig:Evolution} that, if the parameters we consider are extremely close to the Hopf bifurcation, a stable limit cycle attractor originated from supercritical Hopf bifurcation can be created and the trajectory is attracted by the stable limit cycle exhibiting libration. While if the parameter is inside the semicircle formed by the Hopf bifurcation curve in Fig.~\ref{fig:two_parameter_bf}, two stable fixed points $\theta_{0,2}$ and $\theta_{0,4}$ occur and the trajectory will be attracted by one of them without the libration and rotation. 

Finally, it is clear that there are many issues that deserve in-depth exploration in future. For instance, how to extend our model to larger-scale networks and how to analyze the mechanism of synchronization cluster bursting therein are key questions for future work. Even in the three adaptive phase oscillators case, which is the main consideration of this paper, further research is necessary. If synchronization cluster bursting transitions between three or more different synchronization states instead of the two reported in this paper, the question of how to simplify the model and perform detailed analysis becomes even more important, which we intend to discuss in future work.

\begin{acknowledgments}
This work was supported by the Deutsche Forschungsgemeinschaft (DFG, German Research Foundation), Project No. 411803875. MW acknowledges support from China Scholarship Council (CSC) scholarship (Grant No. 202208320297). AA acknowledges support from PIK Werkvertrag 2023-0336. OB acknowledges hospitality at PIK and HU, 
as well as support through the DFG via
project 195170736 - SFB/TRR 109. XH acknowledges support from the National Natural Science Foundation of China (Grant Nos. 12272150 and 12072132).
The authors would like to thank Matheus Rolim Sales for discussions and numerical studies during the initial phase of the project. 
\end{acknowledgments}

\section*{AUTHOR DECLARATIONS}
\subsection*{Conflict of Interest}
The authors have no conflicts to disclose.

\subsection*{Author Contributions}
Mengke Wei: conceptualization, methodology, writing -- original draft. Andreas Amann: methodology, writing -- review \& editing, supervision. Oleksandr Burylko: methodology, writing -- editing. Xiujing Han: supervision.  Serhiy Yanchuk: methodology, writing -- review \& editing, supervision. J\"urgen Kurths: methodology, writing -- review \& editing, supervision.

\section*{Data Availability Statement}

The data that support the findings of this study are available from the corresponding author upon reasonable request.

\appendix
\section{Initial values and parameter values corresponding to Fig.~\ref{fig:R_t_pi}}
To replicate the numerical simulation presented in Fig.~\ref{fig:R_t_pi}, the initial values and parameter settings used are outlined in table \ref{tab:values}. 
\label{Initial values and parameter values}
\begin{table}
\centering
\caption{Initial values and parameter values.}
\label{tab:values}
\begin{tabular}{@{}l@{\hspace{1cm}}l@{\hspace{1cm}}l@{\hspace{1cm}}l@{}} 
\hline
\hline
& \multicolumn{3}{l}{Initial Values} \\

\( \varphi_{1} \) = 6.28121 & \( \kappa_{12} \) = 0.41373 & \( \kappa_{23} \) = -0.15108 & \\
\( \varphi_{2} \) = 3.74488 & \( \kappa_{13} \) = -0.33369 & \( \kappa_{31} \) = 0.08116 & \\
\( \varphi_{3} \) = 3.40165 & \( \kappa_{21} \) = 0.27333 & \( \kappa_{32} \) = 0.02394 & \\

\\
& \multicolumn{3}{l}{Parameter Values} \\

\( \delta_{12} \) = 0.5$\pi$ & \( \delta_{32} \) = 1.5$\pi$  & \( A_{31} \) = 0.3 & \\
\( \delta_{13}\) = 0.88$\pi$  & \( A_{12} \) = 0.1  & \( A_{32} \) = 0.5 & \\
\( \delta_{21} \) = 0.5$\pi$  & \( A_{13} \) = 0.1  & \( \omega_{1} \) = 0.012 & \\
\( \delta_{23} \) = 0.88$\pi$ & \( A_{21} \) =0.2  & \( \omega_{2} \) = 0.007 & \\
\( \delta_{31} \) = 0.5$\pi$ & \( A_{23} \) = 0.3  & \( \omega_{3} \) = 0.003 & \\
\hline
\hline
\end{tabular}
\end{table}

\section{Expressions for parameters in the reduced system}
\label{APPENDIX A}
Based on the phase difference and frequency difference in (\ref{eq: phase difference in 3D system}), it holds that
\begin{equation}
\label{}
\begin{split}
\dot \theta = & \Delta  - \frac{1}{3} \Big[\left({\kappa _{12}} + \frac{{{\kappa _{21}}}}{2}\right)\sin \left(\theta  - \frac{\alpha }{2}\right) \\
&
+ \left({\kappa _{13}} + \frac{{{\kappa _{31}}}}{2}\right)\sin \left(\theta  + \frac{\alpha }{2}\right)
+ \left(\frac{{{\kappa _{32}}}}{2} - \frac{{{\kappa _{23}}}}{2}\right)\sin (\alpha )\Big]\\
{{\dot \kappa }_{12}} =&  - \varepsilon [{\kappa _{12}} + {A_{12}}\sin (\theta  - \frac{\alpha }{2} + {\delta _{12}})]\\
{{\dot \kappa }_{13}} =&  - \varepsilon [{\kappa _{13}} + {A_{13}}\sin (\theta  + \frac{\alpha }{2} + {\delta _{13}})]\\
{{\dot \kappa }_{21}} =&  - \varepsilon [{\kappa _{21}} - {A_{21}}\sin (\theta  - \frac{\alpha }{2} - {\delta _{21}})]\\
{{\dot \kappa }_{23}} =&  - \varepsilon [{\kappa _{23}} + {A_{23}}\sin (\alpha  + {\delta _{23}})]\\
{{\dot \kappa }_{31}} =&  - \varepsilon [{\kappa _{31}} - {A_{31}}\sin (\theta  + \frac{\alpha }{2} - {\delta _{31}})]\\
{{\dot \kappa }_{32}} =&  - \varepsilon [{\kappa _{32}} - {A_{32}}\sin (\alpha  - {\delta _{32}})].
\end{split}
\end{equation}
By letting
\begin{equation}
\label{}
{\kappa _1} = {\kappa _{12}} + \frac{{{\kappa _{21}}}}{2},
    \qquad
{\kappa _2} = {\kappa _{13}} + \frac{{{\kappa _{31}}}}{2},
    \qquad
{\kappa _3} =  - \frac{{{\kappa _{23}}}}{2} + \frac{{{\kappa _{32}}}}{2},
\end{equation}
to combine $\kappa_{ij}$ and $\kappa_{ij}$, the above equation can be further simplified as
\begin{equation}
\label{}
\begin{array}{l}
\dot \theta  = \Delta  - [{\kappa _1}\sin (\theta  - \frac{\alpha }{2}) + {\kappa _2}\sin (\theta  + \frac{\alpha }{2}) + {\kappa _3}\sin (\alpha )]/3\\
{{\dot \kappa }_1} =  - \varepsilon [{\kappa _1} + {A_{12}}\sin (\theta  - \frac{\alpha }{2} + {\delta _{12}}) - \frac{{{A_{21}}}}{2}\sin (\theta  - \frac{\alpha }{2} - {\delta _{21}})]\\
{{\dot \kappa }_2} =  - \varepsilon [{\kappa _2} + {A_{13}}\sin (\theta  + \frac{\alpha }{2} + {\delta _{13}}) - \frac{{{A_{31}}}}{2}\sin (\theta  + \frac{\alpha }{2} - {\delta _{31}})]\\
{{\dot \kappa }_3} =  - \varepsilon [{\kappa _3} - \frac{{{A_{23}}}}{2}\sin (\alpha  + {\delta _{23}}) - \frac{{{A_{32}}}}{2}\sin (\alpha  - {\delta _{32}})].
\end{array}
\end{equation}
Since $\kappa_3$ is a constant, i.e.,
\begin{equation}
\label{eq: kappa _3}
{\kappa _3} = \frac{{{A_{23}}}}{2}\sin (\alpha  + {\delta _{23}}) + \frac{{{A_{32}}}}{2}\sin (\alpha  - {\delta _{32}}),
\end{equation}
we can obtain the simplified three-dimensional system
\begin{equation}
\label{eq: 3-D system}
\begin{split}
\dot \theta  =& \Delta  -\frac{1}{3} \bigg[{\kappa _1}\sin \left(\theta  - \frac{\alpha }{2}\right) + {\kappa _2}\sin \left(\theta  + \frac{\alpha }{2}\right) \\
& + \left(\frac{A_{23}}{2}\sin (\alpha  + {\delta _{23}})
+ \frac{{{A_{32}}}}{2}\sin (\alpha  - {\delta _{32}})\right)\sin (\alpha )\bigg]\\
{{\dot \kappa }_1} =&  - \varepsilon [{\kappa _1} + {A_{12}}\sin (\theta  - \frac{\alpha }{2} + {\delta _{12}}) - \frac{{{A_{21}}}}{2}\sin (\theta  - \frac{\alpha }{2} - {\delta _{21}})]\\
{{\dot \kappa }_2} =&  - \varepsilon [{\kappa _2} + {A_{13}}\sin (\theta  + \frac{\alpha }{2} + {\delta _{13}}) - \frac{{{A_{31}}}}{2}\sin (\theta  + \frac{\alpha }{2} - {\delta _{31}})],
\end{split}
\end{equation}
which can further be reformulated into the form of equation (\ref{eq: simple 3-D system}), where
\begin{widetext}
\begin{align}
\label{}
\beta  &= \left[{A_{23}}\sin (\alpha  + {\delta _{23}}) + {A_{32}}\sin (\alpha  - {\delta _{32}})\right]/2\\
{A_1} &= \sqrt {{{\left( {{A_{12}}\cos ({\delta _{12}}) - 0.5{A_{21}}\cos ({\delta _{21}})} \right)}^2} + {{\left( {{A_{12}}\sin ({\delta _{12}}) + 0.5{A_{21}}\sin ({\delta _{21}})} \right)}^2}}\\ 
{\delta _1} &= \arctan \left( {\frac{{{A_{12}}\sin ({\delta _{12}}) + 0.5{A_{21}}\sin ({\delta _{21}})}}{{{A_{12}}\cos ({\delta _{12}}) - 0.5{A_{21}}\cos ({\delta _{21}})}}} \right)\\
{A_2} &= \sqrt {{{\left( {{A_{13}}\cos ({\delta _{13}}) - 0.5{A_{31}}\cos ({\delta _{31}})} \right)}^2} + {{\left( {{A_{13}}\sin ({\delta _{13}}) + 0.5{A_{31}}\sin ({\delta _{31}})} \right)}^2}} \\
{\delta _2} &= \arctan \left( {\frac{{{A_{13}}\sin ({\delta _{13}}) + 0.5{A_{31}}\sin ({\delta _{31}})}}{{{A_{13}}\cos ({\delta _{13}}) - 0.5{A_{31}}\cos ({\delta _{31}})}}} \right).
\end{align}
\end{widetext}
\section{Derivation of normal form equation}
\label{APPENDIX B}
We can replace the two kappas in system (\ref{eq: simple 3-D system}) with a single complex dynamical quantity
\begin{equation}
    \check{Q}^*=\kappa_{1}e^ {i(-\frac{\alpha}{2})}+\kappa_{2}e^ {i(+\frac{\alpha}{2})}.
\end{equation}
This means that the equation for $\theta$ becomes
\begin{equation}
    \dot{\theta}=\Delta-[\text{Im}\left\{ \check{Q}^* e^ {i\theta}\right\} +\beta\sin(\alpha)]/3. \label{eq: new theta}
\end{equation}
Then the derive equation for $\dot{\check{Q}}$ ($\check{Q}^*$is the complex conjugate of $\check{Q}$) can be given by
\begin{widetext}
\begin{align}
{{\dot {\check{Q}}}} &=  {{\dot \kappa }_1}e^ {i(  \frac{\alpha }{2})} + {{\dot \kappa }_2}e^ {i(-\frac{\alpha }{2})}\\
&= e^ {i( \frac{\alpha }{2})}\left\{ {  -\varepsilon [{\kappa _1} + {A_1}\sin (\theta  - \frac{\alpha }{2} + {\delta _1})]} \right\} + e^ {i(-\frac{\alpha }{2})}\left\{ { - \varepsilon [{\kappa _2} + {A_2}\sin (\theta  + \frac{\alpha }{2} + {\delta _2})]} \right\}\\
&=  - \varepsilon \left[ {{\kappa _1}e^ {i( \frac{\alpha }{2})} + {A_1}e^ {i( \frac{\alpha }{2})}\sin (\theta  - \frac{\alpha }{2} + {\delta _1}) + {\kappa _2}e^ {i(-\frac{\alpha }{2})} + {A_2}e^ {i(-\frac{\alpha }{2})}\sin (\theta  + \frac{\alpha }{2} + {\delta _2})} \right]\\
&=  - \varepsilon \left[ {{\check{Q}} + {A_1}e^ {i( \frac{\alpha }{2})}\sin (\theta  - \frac{\alpha }{2} + {\delta _1}) + {A_2}e^ {i(-\frac{\alpha }{2})}\sin (\theta  + \frac{\alpha }{2} + {\delta _2})} \right].
\end{align}
\end{widetext}
Based on
\begin{equation}
    e^ {ix} = \cos (x) + i\sin (x),
\end{equation}
we can get
\begin{align}
\sin (x) = \frac{{e^ {ix} - e^ {- ix}}}{{2i}}.
\end{align}
For the first term
\begin{equation}
A_1 e^{ i   \left(\frac{\alpha }{2}\right) } \sin \left( \theta  - \frac{\alpha }{2} + \delta_1 \right),
\end{equation}
we can expand it as
\begin{equation}
 \frac{A_1}{2i} \left[ e^ { i \left( \theta + \delta_1  \right) } - e^ {- i \left( \theta + \delta_1 - \alpha\right) } \right].
\end{equation}
For the second term
\begin{equation}
A_2 e^  {i \left( -\frac{\alpha }{2} \right)} \sin \left( \theta  + \frac{\alpha }{2} + \delta_2 \right),
\end{equation}
we can expand it as
\begin{equation}
\frac{A_2}{2i} \left[ e^ {i \left( \theta + \delta_2 \right)} - e^ {-i \left( \theta + \delta_2  + \alpha\right)}  \right].
\end{equation}
Combining the two terms and extracting coefficients associated with $e^{i\theta}$ and $e^{-i\theta}$, we obtain
\begin{equation}
\begin{split}
     &{A_1}e^ {i(  \frac{\alpha }{2})}\sin (\theta  - \frac{\alpha }{2} + {\delta _1}) + {A_2}e^ {i(-\frac{\alpha }{2})}\sin (\theta  + \frac{\alpha }{2} + {\delta _2}) \\
     = &\check{C_1}e^ {i\theta} + \check{C_2}e^ {- i\theta }.
\end{split}
\end{equation}
As a result, the derive equation for $\dot{\check{Q}}$ is
\begin{equation}
    \dot{\check{Q}} = -\epsilon \left[ \check{Q} + \check{C_1} e^{i\theta} + \check{C_2} e^{-i \theta} \right],
\end{equation}
where
\begin{equation}
\begin{split}
\check{C_1} = &
   \frac{1}{{2i}}\left( {A_1}e^{i{\delta _1}} + {A_2}e^{i{\delta _2}} \right)\\
   =&\left(A_{1}\sin\delta_{1}+A_{2}\sin\delta_{2}\right)/2\\
   &- i\left(A_{1}\cos\delta_{1}+A_{2}\cos\delta_{2}\right)/2,\\
\check{C_2} =& 
   - \frac{1}{{2i}}\left[ {A_1}e^{-i\left( {\delta _1 - \alpha } \right)} + {A_2}e^{-i\left( {\delta _2 + \alpha } \right)} \right]\\
   =&\left[A_{1}\sin\left(\delta_{1}-\alpha\right)+A_{2}\sin\left(\delta_{2}+\alpha\right)\right]/2 \\
   &+ i\left[A_{1}\cos\left(\delta_{1}-\alpha\right)+A_{2}\cos\left(\delta_{2}+\alpha\right)\right]/2.
\end{split}
\end{equation}
are complex constants. Furthermore, if we absorb the $\beta$ term into $\left| \Delta  \right|$, we can define:
\begin{equation}
    \check \Delta  = \Delta  - \frac{1}{3}\beta \sin (\alpha ).
\end{equation}
Let's define a new time scale $\tau $ such that
\begin{equation}
    d\tau  = \left| {\check \Delta } \right|dt.
\end{equation}
Then, setting $\hat \varepsilon  = \frac{\epsilon }{{\check \Delta }}$, ${C_1} = \frac{{\check{C_1}}}{{3{{\check \Delta }}}}$, ${C_2} = \frac{{\check{C_2}}}{{3{{\check \Delta }}}}$, the rescaled equation in terms of $\tau $  becomes
\begin{equation}\label{eq: normal form_appendix}
\begin{split}
\dot{\theta} & = 1 - \left| Q \right| \sin\left( \theta - \phi_{Q} \right) \\
\dot{Q} & = -\hat{\epsilon} \left[ Q + C_{1}e^{i\theta} + C_{2}e^{-i\theta} \right],
\end{split}
\end{equation}
where $C_{1},C_{2}\in\mathbb{C}$ and  $\hat{\epsilon}\in\mathbb{R}$ are parameters, and $\theta\in {\mathbf{S}^1}$,  $Q\in\mathbb{C}$ are the dynamical variables, $\phi_{Q}$ represents the argument of $Q$.
\section{Derivation of fixed points of the normal form equation and their stability analysis}
\label{APPENDIX C}
\subsection{Fixed points}
Let us study the fixed points of the normal form (\ref{eq: normal form}) in detail. If $\left(\theta_{0},Q_{0}\right)$ is a fixed point of (\ref{eq: normal form}), then 
\begin{align}
1 & =\text{Im}Q_{0}^{*}e^{i\theta_{0}}\\
-Q_{0} & =C_{1}e^{i\theta_{0}}+C_{2}e^{-i\theta_{0}}.\label{eq:FPQ}
\end{align}
Plugging the second equation into the first gives 
\begin{equation}
\begin{aligned}
1 & =-\text{Im}\left[\left(C_{1}e^{i\theta_{0}}+C_{2}e^{-i\theta_{0}}\right)^{*}e^{i\theta_{0}}\right]\\
 & =-\text{Im}\left[C_{1}^{*}+C_{2}e^{i2\theta_{0}}\right]\\
 & =\text{Im}C_{1}-C_{2}\sin2\theta_{0},
\end{aligned}
\end{equation}
this yields the condition 
\begin{equation}
\sin2\theta_{0}=\frac{\text{Im}C_{1}-1}{C_{2}},\label{eq:thetaFP}
\end{equation}
which can only be fulfilled if 
\begin{equation}
\left|\text{Im}C_{1}-1\right|\leq C_{2}.\label{eq:fpcond}
\end{equation}
As a result, we can obtain four fixed points (\ref{eq: fixed point theta}-\ref{eq: fixed point Q}).

\subsection{Saddle Node bifurcation}
We can plot the area in which (\ref{eq:fpcond}) is fulfilled in a graph with $\text{Im}C_{1}$ on the $x$ axis and $C_{2}$ on the $y$ axis. This is an area above the line  
\begin{equation}
C_{2}=\left|\text{Im}C_{1}-1\right|.
\end{equation}
This line has two parts, one with slope $1$ for $\text{Im}C_{1}>1$, and one with slope $-1$ else.
As we approach this line from above, the four fixed points merge in pairs and vanish as we cross the line.
 Thus this line denotes a saddle node bifurcation.
 More precisely we have for $\text{Im}C_{1}>1$ 
\begin{align}
\theta_{0,1} & \nearrow\frac{\pi}{4}\\
\theta_{0,2} & \searrow\frac{\pi}{4}\\
\theta_{0,3} & \nearrow5\frac{\pi}{4}\\
\theta_{0,4} & \searrow5\frac{\pi}{4}.
\end{align}
At the point where this saddle node bifurcation happens, we have  
\begin{equation}
\begin{aligned}
Q_{0,1} & =-C_{1}e^{i\theta_{0,1}}-C_{2}e^{-i\theta_{0,1}}\\
 & =-C_{1}e^{i\frac{\pi}{4}}-\left(\text{Im}C_{1}-1\right)e^{-i\frac{\pi}{4}}\\
 & =\frac{-C_{1}\left(1+i\right)-\left(\text{Im}C_{1}-1\right)\left(1-i\right)}{\sqrt{2}}\\
 & =\frac{-\left(\text{Re}C_{1}+i\text{Im}C_{1}\right)\left(1+i\right)-\left(\text{Im}C_{1}-1\right)\left(1-i\right)}{\sqrt{2}}\\
 & =\frac{-\left(\text{Re}C_{1}\right)\left(1+i\right)+\left(1-i\right)}{\sqrt{2}}\\
 & =\frac{1-\text{Re}C_{1}-i\left(1+\text{Re}C_{1}\right)}{\sqrt{2}}\\
Q_{0,2} & =-iC_{1}e^{-i\theta_{0,1}}+iC_{2}e^{i\theta_{0,1}}\\
 & =-iC_{1}e^{-i\frac{\pi}{4}}+i\left[\text{Im}C_{1}-1\right]e^{i\frac{\pi}{4}}=Q_{0,1}\\
Q_{0,3} & =-Q_{0,1}\\
Q_{0,4} & =-Q_{0,2}=Q_{0,3}.
\end{aligned}
\end{equation}
Also note that  
\begin{equation}
\left|Q_{0,1}\right|^{2}=1+\left(\text{Re}C_{1}\right)^{2}
\end{equation}
and similar for all other fixed points. This means that the saddle-node bifurcation happens outside of the unit circle. It is therefore different from the naive saddle-node bifurcation in the $\theta$ dynamics, which happens at $\left|Q\right|=1$.

\subsection{Stability of fixed points}
In order to assess the stability of the fixed points, we need to consider the Jacobian matrix of our system. To do this, let us introduce ${\mathop{\rm Re}\nolimits} Q$ and ${\mathop{\rm Im}\nolimits} Q$: 

\begin{align}
Q & ={\mathop{\rm Re}\nolimits} Q+i{\mathop{\rm Im}\nolimits} Q\\
\left|Q\right| & =\sqrt{{\mathop{\rm Re}\nolimits} Q^{2}+{\mathop{\rm Im}\nolimits} Q^{2}}\\
\phi_{Q} & =\arctan\frac{{\mathop{\rm Im}\nolimits} Q}{{\mathop{\rm Re}\nolimits} Q}.
\end{align}
This gives the three-dimensional real system: 

\begin{align}
\dot{\theta} & =1-\left({\mathop{\rm Re}\nolimits} Q\sin\theta-{\mathop{\rm Im}\nolimits} Q\cos\theta\right)\label{eq:thetadot-1}\\
{\mathop{\rm Re}\nolimits} \dot Q & =-\hat{\epsilon}\left[{\mathop{\rm Re}\nolimits} Q+\left(\text{Re}C_{1}+C_{2}\right)\cos\theta-\text{Im}C_{1}\sin\theta\right]\label{eq:Rdot}\\
{\mathop{\rm Im}\nolimits} \dot Q & =-\hat{\epsilon}\left[{\mathop{\rm Im}\nolimits} Q+\text{Im}C_{1}\cos\theta+\left(\text{Re}C_{1}-C_{2}\right)\sin\theta\right].\label{eq:Idot}
\end{align}
We get the Jacobian matrix: 
\begin{equation}
J=\left(\begin{array}{ccc}
-\left({\mathop{\rm Re}\nolimits} Q\cos\theta+{\mathop{\rm Im}\nolimits} Q\sin\theta\right) & -\sin\theta & \cos\theta\\
\hat{\epsilon}\left[\left(\text{Re}C_{1}+C_{2}\right)\sin\theta+\text{Im}C_{1}\cos\theta\right] & -\hat{\epsilon} & 0\\
\hat{\epsilon}\left[\text{Im}C_{1}\sin\theta-\left(\text{Re}C_{1}-C_{2}\right)\cos\theta\right] & 0 & -\hat{\epsilon}
\end{array}\right).
\end{equation}
We are particularly interested in the case of evaluating the Jacobian at a fixed point. At a fixed point we have, using (\ref{eq:Rdot}) and (\ref{eq:Idot}): 
\begin{equation}
\begin{split}
&{\mathop{\rm Re}\nolimits} Q\cos\theta+{\mathop{\rm Im}\nolimits} Q\sin\theta\\ 
 =&-\left[\left(\text{Re}C_{1}+C_{2}\right)\cos\theta
 -\text{Im}C_{1}\sin\theta\right]\cos\theta\\
 &-\left[\text{Im}C_{1}\cos\theta+\left(\text{Re}C_{1}
-C_{2}\right)\sin\theta\right]\sin\theta\\
 =& -\text{Re}C_{1}-C_{2}\cos2\theta,
\end{split}
\end{equation}
and thus 
\begin{equation}\label{eq:Jacobian}
J=\left(\begin{array}{ccc}
\text{Re}C_{1}+C_{2}\cos2\theta & -\sin\theta & \cos\theta\\
\hat{\epsilon}\left[\left(\text{Re}C_{1}+C_{2}\right)\sin\theta+\text{Im}C_{1}\cos\theta\right] & -\hat{\epsilon} & 0\\
\hat{\epsilon}\left[\text{Im}C_{1}\sin\theta-\left(\text{Re}C_{1}-C_{2}\right)\cos\theta\right] & 0 & -\hat{\epsilon}
\end{array}\right).
\end{equation}
We can calculate the characteristic polynomial as
\begin{widetext}
\begin{equation}
\begin{aligned}
\chi\left(\lambda\right) & =\left[\text{Re}C_{1}+C_{2}\cos2\theta-\lambda\right]\left(-\hat{\epsilon}-\lambda\right)^{2}\\
 & -\hat{\epsilon}\left[\left(\text{Re}C_{1}+C_{2}\right)\sin\theta+\text{Im}C_{1}\cos\theta\right]\left[-\sin\theta\right]\left[-\hat{\epsilon}-\lambda\right]\\
 & -\hat{\epsilon}\left[\text{Im}C_{1}\sin\theta-\left(\text{Re}C_{1}-C_{2}\right)\cos\theta\right]\left[-\hat{\epsilon}-\lambda\right]\left[\cos\theta\right]\\
 & =\left[\text{Re}C_{1}+C_{2}\cos2\theta-\lambda\right]\left(\hat{\epsilon}+\lambda\right)^{2}\\
 & +\hat{\epsilon}\left[\hat{\epsilon}+\lambda\right]\left[-\left(\text{Re}C_{1}+C_{2}\right)\sin^{2}\theta-\text{Im}C_{1}\cos\theta\sin\theta+\text{Im}C_{1}\sin\theta\cos\theta-\left(\text{Re}C_{1}-C_{2}\right)\cos^{2}\theta\right]\\
 & =\left[\text{Re}C_{1}+C_{2}\cos2\theta-\lambda\right]\left(\hat{\epsilon}+\lambda\right)^{2}\\
 & +\hat{\epsilon}\left[\hat{\epsilon}+\lambda\right]\left[-\text{Re}C_{1}+C_{2}\left(\cos^{2}\theta-\sin^{2}\theta\right)\right]\\
 & =\left(\hat{\epsilon}+\lambda\right)\left\{ \left[\text{Re}C_{1}+C_{2}\cos2\theta-\lambda\right]\left(\hat{\epsilon}+\lambda\right)+\hat{\epsilon}\left[-\text{Re}C_{1}+C_{2}\cos2\theta\right]\right\} \\
 & =\left(\hat{\epsilon}+\lambda\right)\left\{ \text{Re}C_{1}\lambda+C_{2}\cos2\theta\left(\lambda+2\hat{\epsilon}\right)-\lambda^{2}-\lambda\hat{\epsilon}\right\} \\
 & =-\left(\hat{\epsilon}+\lambda\right)\left\{ \lambda^{2}-\lambda\left[\text{Re}C_{1}+C_{2}\cos2\theta-\hat{\epsilon}\right]-2\hat{\epsilon} C_{2}\cos2\theta\right\}.
\end{aligned}
\end{equation}    
\end{widetext}
This means that one eigenvalue is always $\lambda_{1}=-\hat{\epsilon}$, which corresponds to a stable eigen-direction. In addition, we have two other eigenvalues, which we get as solutions of the quadratic equation
\begin{align}
\lambda^{2}-\lambda T+D & =0\\
T & =\text{Re}C_{1}+C_{2}\cos2\theta-\hat{\epsilon}\\
D & =-2\hat{\epsilon} C_{2}\cos2\theta\\
\lambda_{2/3} & =\frac{T}{2}\pm\sqrt{\left(\frac{T}{2}\right)^{2}-D}.
\end{align}
We see that for $\theta_{0,1}$
\begin{align}
\cos2\theta_{0,1} & \geq0,\\
D & \leq0,
\end{align}
therefore $\theta_{0,1}$  is a saddle with two unstable and one stable direction. Similarly $\theta_{0,3}$ is also a saddle. On the other hand for $\theta_{0,2}$ we have
\begin{align}
\cos2\theta_{0,2} & \leq0,\\
D & \geq0.
\end{align}
The stability therefore depends on the sign of $T$. If $T<0$, we have a stable node or focus with three stable directions; if $T>0$, we have two unstable directions. At $T=0$, we have a Hopf bifurcation. Let us express the condition for the Hopf in the original parameters of the system. At the fixed point we have
\begin{align}
C_{2}\cos2\theta_{0} & =\pm C_{2} \sqrt{1-\sin^{2}2\theta_{0}}=\pm\sqrt{C_{2}^{2}-\left(\text{Im}C_{1}-1\right)^{2}}
\end{align}
through (\ref{eq:thetaFP}).
Thus the Hopf condition is given by 
\begin{equation}
\text{Re}C_{1}-\hat{\epsilon}=\mp\sqrt{C_{2}^{2}-\left(\text{Im}C_{1}-1\right)^{2}}
\end{equation}
or more elegantly  
\begin{equation}
\left|C_{1}-(\hat{\epsilon}+i)\right|=C_{2}.
\end{equation}

\subsection{Eigen-directions of the Jacobian matrix}
The Jacobian matrix (\ref{eq:Jacobian}) is remarkable in that at the fixed points it only depends on the dynamical variable $\theta$. We verify that
\begin{equation}\left.J\left(\begin{array}{c}0\\\cos\theta\\\sin\theta\end{array}\right.\right)=-\epsilon\left(\begin{array}{c}0\\\cos\theta\\\sin\theta\end{array}\right),\end{equation}
and therefore the (slowly) stable eigen-direction with which we approach the fixed point is given by
\begin{equation}\left.v=\left(\begin{array}{c}0\\\cos\theta\\\sin\theta\end{array}\right.\right).\end{equation}


\begin{thebibliography}{58}%
\makeatletter
\providecommand \@ifxundefined [1]{%
 \@ifx{#1\undefined}
}%
\providecommand \@ifnum [1]{%
 \ifnum #1\expandafter \@firstoftwo
 \else \expandafter \@secondoftwo
 \fi
}%
\providecommand \@ifx [1]{%
 \ifx #1\expandafter \@firstoftwo
 \else \expandafter \@secondoftwo
 \fi
}%
\providecommand \natexlab [1]{#1}%
\providecommand \enquote  [1]{``#1''}%
\providecommand \bibnamefont  [1]{#1}%
\providecommand \bibfnamefont [1]{#1}%
\providecommand \citenamefont [1]{#1}%
\providecommand \href@noop [0]{\@secondoftwo}%
\providecommand \href [0]{\begingroup \@sanitize@url \@href}%
\providecommand \@href[1]{\@@startlink{#1}\@@href}%
\providecommand \@@href[1]{\endgroup#1\@@endlink}%
\providecommand \@sanitize@url [0]{\catcode `\\12\catcode `\$12\catcode
  `\&12\catcode `\#12\catcode `\^12\catcode `\_12\catcode `\%12\relax}%
\providecommand \@@startlink[1]{}%
\providecommand \@@endlink[0]{}%
\providecommand \url  [0]{\begingroup\@sanitize@url \@url }%
\providecommand \@url [1]{\endgroup\@href {#1}{\urlprefix }}%
\providecommand \urlprefix  [0]{URL }%
\providecommand \Eprint [0]{\href }%
\providecommand \doibase [0]{http://dx.doi.org/}%
\providecommand \selectlanguage [0]{\@gobble}%
\providecommand \bibinfo  [0]{\@secondoftwo}%
\providecommand \bibfield  [0]{\@secondoftwo}%
\providecommand \translation [1]{[#1]}%
\providecommand \BibitemOpen [0]{}%
\providecommand \bibitemStop [0]{}%
\providecommand \bibitemNoStop [0]{.\EOS\space}%
\providecommand \EOS [0]{\spacefactor3000\relax}%
\providecommand \BibitemShut  [1]{\csname bibitem#1\endcsname}%
\let\auto@bib@innerbib\@empty
\bibitem [{\citenamefont {Christensen}\ \emph {et~al.}(1998)\citenamefont
  {Christensen}, \citenamefont {Donangelo}, \citenamefont {Koiller},\ and\
  \citenamefont {Sneppen}}]{christensen1998evolution}%
  \BibitemOpen
  \bibfield  {author} {\bibinfo {author} {\bibfnamefont {K.}~\bibnamefont
  {Christensen}}, \bibinfo {author} {\bibfnamefont {R.}~\bibnamefont
  {Donangelo}}, \bibinfo {author} {\bibfnamefont {B.}~\bibnamefont {Koiller}},
  \ and\ \bibinfo {author} {\bibfnamefont {K.}~\bibnamefont {Sneppen}},\
  }\bibfield  {title} {\enquote {\bibinfo {title} {Evolution of random
  networks},}\ }\href@noop {} {\bibfield  {journal} {\bibinfo  {journal}
  {Physical Review Letters}\ }\textbf {\bibinfo {volume} {81}},\ \bibinfo
  {pages} {2380} (\bibinfo {year} {1998})}\BibitemShut {NoStop}%
\bibitem [{\citenamefont {Bornholdt}\ and\ \citenamefont
  {Rohlf}(2000)}]{bornholdt2000topological}%
  \BibitemOpen
  \bibfield  {author} {\bibinfo {author} {\bibfnamefont {S.}~\bibnamefont
  {Bornholdt}}\ and\ \bibinfo {author} {\bibfnamefont {T.}~\bibnamefont
  {Rohlf}},\ }\bibfield  {title} {\enquote {\bibinfo {title} {Topological
  evolution of dynamical networks: Global criticality from local dynamics},}\
  }\href@noop {} {\bibfield  {journal} {\bibinfo  {journal} {Physical Review
  Letters}\ }\textbf {\bibinfo {volume} {84}},\ \bibinfo {pages} {6114}
  (\bibinfo {year} {2000})}\BibitemShut {NoStop}%
\bibitem [{\citenamefont {Gross}\ and\ \citenamefont
  {Blasius}(2008)}]{gross2008adaptive}%
  \BibitemOpen
  \bibfield  {author} {\bibinfo {author} {\bibfnamefont {T.}~\bibnamefont
  {Gross}}\ and\ \bibinfo {author} {\bibfnamefont {B.}~\bibnamefont
  {Blasius}},\ }\bibfield  {title} {\enquote {\bibinfo {title} {Adaptive
  coevolutionary networks: A review},}\ }\href {\doibase
  10.1098/rsif.2007.1229} {\bibfield  {journal} {\bibinfo  {journal} {Journal
  of The Royal Society Interface}\ }\textbf {\bibinfo {volume} {5}},\ \bibinfo
  {pages} {259--271} (\bibinfo {year} {2008})}\BibitemShut {NoStop}%
\bibitem [{\citenamefont {Berner}\ \emph {et~al.}(2023)\citenamefont {Berner},
  \citenamefont {Gross}, \citenamefont {Kuehn}, \citenamefont {Kurths},\ and\
  \citenamefont {Yanchuk}}]{Berner2023}%
  \BibitemOpen
  \bibfield  {author} {\bibinfo {author} {\bibfnamefont {R.}~\bibnamefont
  {Berner}}, \bibinfo {author} {\bibfnamefont {T.}~\bibnamefont {Gross}},
  \bibinfo {author} {\bibfnamefont {C.}~\bibnamefont {Kuehn}}, \bibinfo
  {author} {\bibfnamefont {J.}~\bibnamefont {Kurths}}, \ and\ \bibinfo {author}
  {\bibfnamefont {S.}~\bibnamefont {Yanchuk}},\ }\bibfield  {title} {\enquote
  {\bibinfo {title} {Adaptive dynamical networks},}\ }\href {\doibase
  10.1016/j.physrep.2023.08.001} {\bibfield  {journal} {\bibinfo  {journal}
  {Physics Reports}\ }\textbf {\bibinfo {volume} {1031}},\ \bibinfo {pages}
  {1--59} (\bibinfo {year} {2023})}\BibitemShut {NoStop}%
\bibitem [{\citenamefont {Sawicki}\ \emph {et~al.}(2023)\citenamefont
  {Sawicki}, \citenamefont {Berner}, \citenamefont {Loos}, \citenamefont
  {Anvari}, \citenamefont {Bader}, \citenamefont {Barfuss}, \citenamefont
  {Botta}, \citenamefont {Brede}, \citenamefont {Franovi{\'c}}, \citenamefont
  {Gauthier} \emph {et~al.}}]{Sawicki2023}%
  \BibitemOpen
  \bibfield  {author} {\bibinfo {author} {\bibfnamefont {J.}~\bibnamefont
  {Sawicki}}, \bibinfo {author} {\bibfnamefont {R.}~\bibnamefont {Berner}},
  \bibinfo {author} {\bibfnamefont {S.~A.}\ \bibnamefont {Loos}}, \bibinfo
  {author} {\bibfnamefont {M.}~\bibnamefont {Anvari}}, \bibinfo {author}
  {\bibfnamefont {R.}~\bibnamefont {Bader}}, \bibinfo {author} {\bibfnamefont
  {W.}~\bibnamefont {Barfuss}}, \bibinfo {author} {\bibfnamefont
  {N.}~\bibnamefont {Botta}}, \bibinfo {author} {\bibfnamefont
  {N.}~\bibnamefont {Brede}}, \bibinfo {author} {\bibfnamefont
  {I.}~\bibnamefont {Franovi{\'c}}}, \bibinfo {author} {\bibfnamefont {D.~J.}\
  \bibnamefont {Gauthier}},  \emph {et~al.},\ }\bibfield  {title} {\enquote
  {\bibinfo {title} {Perspectives on adaptive dynamical systems},}\ }\href@noop
  {} {\bibfield  {journal} {\bibinfo  {journal} {Chaos: An Interdisciplinary
  Journal of Nonlinear Science}\ }\textbf {\bibinfo {volume} {33}} (\bibinfo
  {year} {2023})}\BibitemShut {NoStop}%
\bibitem [{\citenamefont {Sayama}\ \emph {et~al.}(2013)\citenamefont {Sayama},
  \citenamefont {Pestov}, \citenamefont {Schmidt}, \citenamefont {Bush},
  \citenamefont {Wong}, \citenamefont {Yamanoi},\ and\ \citenamefont
  {Gross}}]{sayama2013modeling}%
  \BibitemOpen
  \bibfield  {author} {\bibinfo {author} {\bibfnamefont {H.}~\bibnamefont
  {Sayama}}, \bibinfo {author} {\bibfnamefont {I.}~\bibnamefont {Pestov}},
  \bibinfo {author} {\bibfnamefont {J.}~\bibnamefont {Schmidt}}, \bibinfo
  {author} {\bibfnamefont {B.~J.}\ \bibnamefont {Bush}}, \bibinfo {author}
  {\bibfnamefont {C.}~\bibnamefont {Wong}}, \bibinfo {author} {\bibfnamefont
  {J.}~\bibnamefont {Yamanoi}}, \ and\ \bibinfo {author} {\bibfnamefont
  {T.}~\bibnamefont {Gross}},\ }\bibfield  {title} {\enquote {\bibinfo {title}
  {Modeling complex systems with adaptive networks},}\ }\href@noop {}
  {\bibfield  {journal} {\bibinfo  {journal} {Computers \& Mathematics with
  Applications}\ }\textbf {\bibinfo {volume} {65}},\ \bibinfo {pages}
  {1645--1664} (\bibinfo {year} {2013})}\BibitemShut {NoStop}%
\bibitem [{\citenamefont {Hebb}(1949)}]{hebb1949organization}%
  \BibitemOpen
  \bibfield  {author} {\bibinfo {author} {\bibfnamefont {D.}~\bibnamefont
  {Hebb}},\ }\href@noop {} {\emph {\bibinfo {title} {The organization of
  behavior. A neuropsychological theory}}}\ (\bibinfo  {publisher} {John
  Wiley},\ \bibinfo {year} {1949})\BibitemShut {NoStop}%
\bibitem [{\citenamefont {Den{\`e}ve}, \citenamefont {Alemi},\ and\
  \citenamefont {Bourdoukan}(2017)}]{deneve2017brain}%
  \BibitemOpen
  \bibfield  {author} {\bibinfo {author} {\bibfnamefont {S.}~\bibnamefont
  {Den{\`e}ve}}, \bibinfo {author} {\bibfnamefont {A.}~\bibnamefont {Alemi}}, \
  and\ \bibinfo {author} {\bibfnamefont {R.}~\bibnamefont {Bourdoukan}},\
  }\bibfield  {title} {\enquote {\bibinfo {title} {The brain as an efficient
  and robust adaptive learner},}\ }\href@noop {} {\bibfield  {journal}
  {\bibinfo  {journal} {Neuron}\ }\textbf {\bibinfo {volume} {94}},\ \bibinfo
  {pages} {969--977} (\bibinfo {year} {2017})}\BibitemShut {NoStop}%
\bibitem [{\citenamefont {Almaatouq}\ \emph {et~al.}(2020)\citenamefont
  {Almaatouq}, \citenamefont {Noriega-Campero}, \citenamefont {Alotaibi},
  \citenamefont {Krafft}, \citenamefont {Moussaid},\ and\ \citenamefont
  {Pentland}}]{almaatouq2020adaptive}%
  \BibitemOpen
  \bibfield  {author} {\bibinfo {author} {\bibfnamefont {A.}~\bibnamefont
  {Almaatouq}}, \bibinfo {author} {\bibfnamefont {A.}~\bibnamefont
  {Noriega-Campero}}, \bibinfo {author} {\bibfnamefont {A.}~\bibnamefont
  {Alotaibi}}, \bibinfo {author} {\bibfnamefont {P.}~\bibnamefont {Krafft}},
  \bibinfo {author} {\bibfnamefont {M.}~\bibnamefont {Moussaid}}, \ and\
  \bibinfo {author} {\bibfnamefont {A.}~\bibnamefont {Pentland}},\ }\bibfield
  {title} {\enquote {\bibinfo {title} {Adaptive social networks promote the
  wisdom of crowds},}\ }\href@noop {} {\bibfield  {journal} {\bibinfo
  {journal} {Proceedings of the National Academy of Sciences}\ }\textbf
  {\bibinfo {volume} {117}},\ \bibinfo {pages} {11379--11386} (\bibinfo {year}
  {2020})}\BibitemShut {NoStop}%
\bibitem [{\citenamefont {Landi}\ \emph {et~al.}(2018)\citenamefont {Landi},
  \citenamefont {Minoarivelo}, \citenamefont {Br{\"a}nnstr{\"o}m},
  \citenamefont {Hui},\ and\ \citenamefont {Dieckmann}}]{landi2018complexity}%
  \BibitemOpen
  \bibfield  {author} {\bibinfo {author} {\bibfnamefont {P.}~\bibnamefont
  {Landi}}, \bibinfo {author} {\bibfnamefont {H.~O.}\ \bibnamefont
  {Minoarivelo}}, \bibinfo {author} {\bibfnamefont {{\AA}.}~\bibnamefont
  {Br{\"a}nnstr{\"o}m}}, \bibinfo {author} {\bibfnamefont {C.}~\bibnamefont
  {Hui}}, \ and\ \bibinfo {author} {\bibfnamefont {U.}~\bibnamefont
  {Dieckmann}},\ }\bibfield  {title} {\enquote {\bibinfo {title} {Complexity
  and stability of ecological networks: a review of the theory},}\ }\href@noop
  {} {\bibfield  {journal} {\bibinfo  {journal} {Population Ecology}\ }\textbf
  {\bibinfo {volume} {60}},\ \bibinfo {pages} {319--345} (\bibinfo {year}
  {2018})}\BibitemShut {NoStop}%
\bibitem [{\citenamefont {Raimundo}, \citenamefont {Guimar{\~a}es},\ and\
  \citenamefont {Evans}(2018)}]{raimundoAdaptiveNetworksRestoration2018}%
  \BibitemOpen
  \bibfield  {author} {\bibinfo {author} {\bibfnamefont {R.~L.}\ \bibnamefont
  {Raimundo}}, \bibinfo {author} {\bibfnamefont {P.~R.}\ \bibnamefont
  {Guimar{\~a}es}}, \ and\ \bibinfo {author} {\bibfnamefont {D.~M.}\
  \bibnamefont {Evans}},\ }\bibfield  {title} {\enquote {\bibinfo {title}
  {Adaptive {{Networks}} for {{Restoration Ecology}}},}\ }\href {\doibase
  10.1016/j.tree.2018.06.002} {\bibfield  {journal} {\bibinfo  {journal}
  {Trends in Ecology and Evolution}\ }\textbf {\bibinfo {volume} {33}},\
  \bibinfo {pages} {664--675} (\bibinfo {year} {2018})}\BibitemShut {NoStop}%
\bibitem [{\citenamefont {Maistrenko}\ \emph {et~al.}(2007)\citenamefont
  {Maistrenko}, \citenamefont {Lysyansky}, \citenamefont {Hauptmann},
  \citenamefont {Burylko},\ and\ \citenamefont
  {Tass}}]{maistrenko2007multistability}%
  \BibitemOpen
  \bibfield  {author} {\bibinfo {author} {\bibfnamefont {Y.~L.}\ \bibnamefont
  {Maistrenko}}, \bibinfo {author} {\bibfnamefont {B.}~\bibnamefont
  {Lysyansky}}, \bibinfo {author} {\bibfnamefont {C.}~\bibnamefont
  {Hauptmann}}, \bibinfo {author} {\bibfnamefont {O.}~\bibnamefont {Burylko}},
  \ and\ \bibinfo {author} {\bibfnamefont {P.~A.}\ \bibnamefont {Tass}},\
  }\bibfield  {title} {\enquote {\bibinfo {title} {Multistability in the
  {K}uramoto model with synaptic plasticity},}\ }\href@noop {} {\bibfield
  {journal} {\bibinfo  {journal} {Physical Review E}\ }\textbf {\bibinfo
  {volume} {75}},\ \bibinfo {pages} {066207} (\bibinfo {year}
  {2007})}\BibitemShut {NoStop}%
\bibitem [{\citenamefont {Guti{\'e}rrez}\ \emph {et~al.}(2011)\citenamefont
  {Guti{\'e}rrez}, \citenamefont {Amann}, \citenamefont {Assenza},
  \citenamefont {G{\'o}mez-Gardenes}, \citenamefont {Latora},\ and\
  \citenamefont {Boccaletti}}]{gutierrez2011emerging}%
  \BibitemOpen
  \bibfield  {author} {\bibinfo {author} {\bibfnamefont {R.}~\bibnamefont
  {Guti{\'e}rrez}}, \bibinfo {author} {\bibfnamefont {A.}~\bibnamefont
  {Amann}}, \bibinfo {author} {\bibfnamefont {S.}~\bibnamefont {Assenza}},
  \bibinfo {author} {\bibfnamefont {J.}~\bibnamefont {G{\'o}mez-Gardenes}},
  \bibinfo {author} {\bibfnamefont {V.}~\bibnamefont {Latora}}, \ and\ \bibinfo
  {author} {\bibfnamefont {S.}~\bibnamefont {Boccaletti}},\ }\bibfield  {title}
  {\enquote {\bibinfo {title} {Emerging meso-and macroscales from
  synchronization of adaptive networks},}\ }\href@noop {} {\bibfield  {journal}
  {\bibinfo  {journal} {Physical review letters}\ }\textbf {\bibinfo {volume}
  {107}},\ \bibinfo {pages} {234103} (\bibinfo {year} {2011})}\BibitemShut
  {NoStop}%
\bibitem [{\citenamefont {Burylko}, \citenamefont {Kazanovich},\ and\
  \citenamefont {Borisyuk}(2018)}]{Burylko2018Winner}%
  \BibitemOpen
  \bibfield  {author} {\bibinfo {author} {\bibfnamefont {O.}~\bibnamefont
  {Burylko}}, \bibinfo {author} {\bibfnamefont {Y.}~\bibnamefont {Kazanovich}},
  \ and\ \bibinfo {author} {\bibfnamefont {R.}~\bibnamefont {Borisyuk}},\
  }\bibfield  {title} {\enquote {\bibinfo {title} {Winner-take-all in a phase
  oscillator system with adaptation},}\ }\href
  {https://doi.org/10.1038s41598-017-18666-3} {\bibfield  {journal} {\bibinfo
  {journal} {Scientific Reports}\ }\textbf {\bibinfo {volume} {8}},\ \bibinfo
  {pages} {416} (\bibinfo {year} {2018})}\BibitemShut {NoStop}%
\bibitem [{\citenamefont {Keane}\ \emph {et~al.}(2023)\citenamefont {Keane},
  \citenamefont {Neff}, \citenamefont {Blaha}, \citenamefont {Amann},\ and\
  \citenamefont {H{\"o}vel}}]{keane2023transitional}%
  \BibitemOpen
  \bibfield  {author} {\bibinfo {author} {\bibfnamefont {A.}~\bibnamefont
  {Keane}}, \bibinfo {author} {\bibfnamefont {A.}~\bibnamefont {Neff}},
  \bibinfo {author} {\bibfnamefont {K.}~\bibnamefont {Blaha}}, \bibinfo
  {author} {\bibfnamefont {A.}~\bibnamefont {Amann}}, \ and\ \bibinfo {author}
  {\bibfnamefont {P.}~\bibnamefont {H{\"o}vel}},\ }\bibfield  {title} {\enquote
  {\bibinfo {title} {Transitional cluster dynamics in a model for delay-coupled
  chemical oscillators},}\ }\href@noop {} {\bibfield  {journal} {\bibinfo
  {journal} {Chaos: An Interdisciplinary Journal of Nonlinear Science}\
  }\textbf {\bibinfo {volume} {33}} (\bibinfo {year} {2023})}\BibitemShut
  {NoStop}%
\bibitem [{\citenamefont {Burylko}\ \emph {et~al.}(2018)\citenamefont
  {Burylko}, \citenamefont {Mielke}, \citenamefont {Wolfrum},\ and\
  \citenamefont {Yanchuk}}]{burylko2018coexistence}%
  \BibitemOpen
  \bibfield  {author} {\bibinfo {author} {\bibfnamefont {O.}~\bibnamefont
  {Burylko}}, \bibinfo {author} {\bibfnamefont {A.}~\bibnamefont {Mielke}},
  \bibinfo {author} {\bibfnamefont {M.}~\bibnamefont {Wolfrum}}, \ and\
  \bibinfo {author} {\bibfnamefont {S.}~\bibnamefont {Yanchuk}},\ }\bibfield
  {title} {\enquote {\bibinfo {title} {Coexistence of {H}amiltonian-like and
  dissipative dynamics in rings of coupled phase oscillators with
  skew-symmetric coupling},}\ }\href@noop {} {\bibfield  {journal} {\bibinfo
  {journal} {SIAM Journal on Applied Dynamical Systems}\ }\textbf {\bibinfo
  {volume} {17}},\ \bibinfo {pages} {2076--2105} (\bibinfo {year}
  {2018})}\BibitemShut {NoStop}%
\bibitem [{\citenamefont {D{\"o}rfler}\ and\ \citenamefont
  {Bullo}(2014)}]{reviews3}%
  \BibitemOpen
  \bibfield  {author} {\bibinfo {author} {\bibfnamefont {F.}~\bibnamefont
  {D{\"o}rfler}}\ and\ \bibinfo {author} {\bibfnamefont {F.}~\bibnamefont
  {Bullo}},\ }\bibfield  {title} {\enquote {\bibinfo {title} {Synchronization
  in complex networks of phase oscillators: A survey},}\ }\href@noop {}
  {\bibfield  {journal} {\bibinfo  {journal} {Automatica}\ }\textbf {\bibinfo
  {volume} {50}},\ \bibinfo {pages} {1539--1564} (\bibinfo {year}
  {2014})}\BibitemShut {NoStop}%
\bibitem [{\citenamefont {Kasatkin}\ \emph
  {et~al.}(2017{\natexlab{a}})\citenamefont {Kasatkin}, \citenamefont
  {Yanchuk}, \citenamefont {Sch{\"o}ll},\ and\ \citenamefont
  {Nekorkin}}]{physics3}%
  \BibitemOpen
  \bibfield  {author} {\bibinfo {author} {\bibfnamefont {D.}~\bibnamefont
  {Kasatkin}}, \bibinfo {author} {\bibfnamefont {S.}~\bibnamefont {Yanchuk}},
  \bibinfo {author} {\bibfnamefont {E.}~\bibnamefont {Sch{\"o}ll}}, \ and\
  \bibinfo {author} {\bibfnamefont {V.}~\bibnamefont {Nekorkin}},\ }\bibfield
  {title} {\enquote {\bibinfo {title} {Self-organized emergence of multilayer
  structure and chimera states in dynamical networks with adaptive
  couplings},}\ }\href@noop {} {\bibfield  {journal} {\bibinfo  {journal}
  {Physical Review E}\ }\textbf {\bibinfo {volume} {96}},\ \bibinfo {pages}
  {062211} (\bibinfo {year} {2017}{\natexlab{a}})}\BibitemShut {NoStop}%
\bibitem [{\citenamefont {Berner}, \citenamefont {Scholl},\ and\ \citenamefont
  {Yanchuk}(2019)}]{berner2019multiclusters}%
  \BibitemOpen
  \bibfield  {author} {\bibinfo {author} {\bibfnamefont {R.}~\bibnamefont
  {Berner}}, \bibinfo {author} {\bibfnamefont {E.}~\bibnamefont {Scholl}}, \
  and\ \bibinfo {author} {\bibfnamefont {S.}~\bibnamefont {Yanchuk}},\
  }\bibfield  {title} {\enquote {\bibinfo {title} {Multiclusters in networks of
  adaptively coupled phase oscillators},}\ }\href@noop {} {\bibfield  {journal}
  {\bibinfo  {journal} {SIAM Journal on Applied Dynamical Systems}\ }\textbf
  {\bibinfo {volume} {18}},\ \bibinfo {pages} {2227--2266} (\bibinfo {year}
  {2019})}\BibitemShut {NoStop}%
\bibitem [{\citenamefont {Fialkowski}\ \emph {et~al.}(2023)\citenamefont
  {Fialkowski}, \citenamefont {Yanchuk}, \citenamefont {Sokolov}, \citenamefont
  {Sch{\"o}ll}, \citenamefont {Gottwald},\ and\ \citenamefont
  {Berner}}]{FIA23}%
  \BibitemOpen
  \bibfield  {author} {\bibinfo {author} {\bibfnamefont {J.}~\bibnamefont
  {Fialkowski}}, \bibinfo {author} {\bibfnamefont {S.}~\bibnamefont {Yanchuk}},
  \bibinfo {author} {\bibfnamefont {I.~M.}\ \bibnamefont {Sokolov}}, \bibinfo
  {author} {\bibfnamefont {E.}~\bibnamefont {Sch{\"o}ll}}, \bibinfo {author}
  {\bibfnamefont {G.~A.}\ \bibnamefont {Gottwald}}, \ and\ \bibinfo {author}
  {\bibfnamefont {R.}~\bibnamefont {Berner}},\ }\bibfield  {title} {\enquote
  {\bibinfo {title} {Heterogeneous nucleation in finite-size adaptive dynamical
  networks},}\ }\href@noop {} {\bibfield  {journal} {\bibinfo  {journal} {Phys.
  Rev. Lett.}\ }\textbf {\bibinfo {volume} {130}},\ \bibinfo {pages} {067402}
  (\bibinfo {year} {2023})}\BibitemShut {NoStop}%
\bibitem [{\citenamefont {Cross}\ \emph {et~al.}(2004)\citenamefont {Cross},
  \citenamefont {Zumdieck}, \citenamefont {Lifshitz},\ and\ \citenamefont
  {Rogers}}]{physics1}%
  \BibitemOpen
  \bibfield  {author} {\bibinfo {author} {\bibfnamefont {M.}~\bibnamefont
  {Cross}}, \bibinfo {author} {\bibfnamefont {A.}~\bibnamefont {Zumdieck}},
  \bibinfo {author} {\bibfnamefont {R.}~\bibnamefont {Lifshitz}}, \ and\
  \bibinfo {author} {\bibfnamefont {J.}~\bibnamefont {Rogers}},\ }\bibfield
  {title} {\enquote {\bibinfo {title} {Synchronization by nonlinear frequency
  pulling},}\ }\href@noop {} {\bibfield  {journal} {\bibinfo  {journal}
  {Physical Review Letters}\ }\textbf {\bibinfo {volume} {93}},\ \bibinfo
  {pages} {224101} (\bibinfo {year} {2004})}\BibitemShut {NoStop}%
\bibitem [{\citenamefont {Maistrenko}\ \emph {et~al.}(2004)\citenamefont
  {Maistrenko}, \citenamefont {Popovych}, \citenamefont {Burylko},\ and\
  \citenamefont {Tass}}]{physics2}%
  \BibitemOpen
  \bibfield  {author} {\bibinfo {author} {\bibfnamefont {Y.}~\bibnamefont
  {Maistrenko}}, \bibinfo {author} {\bibfnamefont {O.}~\bibnamefont
  {Popovych}}, \bibinfo {author} {\bibfnamefont {O.}~\bibnamefont {Burylko}}, \
  and\ \bibinfo {author} {\bibfnamefont {P.~A.}\ \bibnamefont {Tass}},\
  }\bibfield  {title} {\enquote {\bibinfo {title} {Mechanism of
  desynchronization in the finite-dimensional {K}uramoto model},}\ }\href@noop
  {} {\bibfield  {journal} {\bibinfo  {journal} {Physical Review Letters}\
  }\textbf {\bibinfo {volume} {93}},\ \bibinfo {pages} {084102} (\bibinfo
  {year} {2004})}\BibitemShut {NoStop}%
\bibitem [{\citenamefont {Pal}\ \emph {et~al.}(2017)\citenamefont {Pal},
  \citenamefont {Tradonsky}, \citenamefont {Chriki}, \citenamefont {Friesem},\
  and\ \citenamefont {Davidson}}]{physics4}%
  \BibitemOpen
  \bibfield  {author} {\bibinfo {author} {\bibfnamefont {V.}~\bibnamefont
  {Pal}}, \bibinfo {author} {\bibfnamefont {C.}~\bibnamefont {Tradonsky}},
  \bibinfo {author} {\bibfnamefont {R.}~\bibnamefont {Chriki}}, \bibinfo
  {author} {\bibfnamefont {A.~A.}\ \bibnamefont {Friesem}}, \ and\ \bibinfo
  {author} {\bibfnamefont {N.}~\bibnamefont {Davidson}},\ }\bibfield  {title}
  {\enquote {\bibinfo {title} {Observing dissipative topological defects with
  coupled lasers},}\ }\href@noop {} {\bibfield  {journal} {\bibinfo  {journal}
  {Physical Review Letters}\ }\textbf {\bibinfo {volume} {119}},\ \bibinfo
  {pages} {013902} (\bibinfo {year} {2017})}\BibitemShut {NoStop}%
\bibitem [{\citenamefont {Yoshimoto}, \citenamefont {Yoshikawa},\ and\
  \citenamefont {Mori}(1993)}]{chemicaloscillators1}%
  \BibitemOpen
  \bibfield  {author} {\bibinfo {author} {\bibfnamefont {M.}~\bibnamefont
  {Yoshimoto}}, \bibinfo {author} {\bibfnamefont {K.}~\bibnamefont
  {Yoshikawa}}, \ and\ \bibinfo {author} {\bibfnamefont {Y.}~\bibnamefont
  {Mori}},\ }\bibfield  {title} {\enquote {\bibinfo {title} {Coupling among
  three chemical oscillators: synchronization, phase death, and frustration},}\
  }\href@noop {} {\bibfield  {journal} {\bibinfo  {journal} {Physical Review
  E}\ }\textbf {\bibinfo {volume} {47}},\ \bibinfo {pages} {864} (\bibinfo
  {year} {1993})}\BibitemShut {NoStop}%
\bibitem [{\citenamefont {Kiss}, \citenamefont {Zhai},\ and\ \citenamefont
  {Hudson}(2002)}]{chemicaloscillators2}%
  \BibitemOpen
  \bibfield  {author} {\bibinfo {author} {\bibfnamefont {I.~Z.}\ \bibnamefont
  {Kiss}}, \bibinfo {author} {\bibfnamefont {Y.}~\bibnamefont {Zhai}}, \ and\
  \bibinfo {author} {\bibfnamefont {J.~L.}\ \bibnamefont {Hudson}},\ }\bibfield
   {title} {\enquote {\bibinfo {title} {Emerging coherence in a population of
  chemical oscillators},}\ }\href@noop {} {\bibfield  {journal} {\bibinfo
  {journal} {Science}\ }\textbf {\bibinfo {volume} {296}},\ \bibinfo {pages}
  {1676--1678} (\bibinfo {year} {2002})}\BibitemShut {NoStop}%
\bibitem [{\citenamefont {Rohden}\ \emph {et~al.}(2012)\citenamefont {Rohden},
  \citenamefont {Sorge}, \citenamefont {Timme},\ and\ \citenamefont
  {Witthaut}}]{electrical_engineering1}%
  \BibitemOpen
  \bibfield  {author} {\bibinfo {author} {\bibfnamefont {M.}~\bibnamefont
  {Rohden}}, \bibinfo {author} {\bibfnamefont {A.}~\bibnamefont {Sorge}},
  \bibinfo {author} {\bibfnamefont {M.}~\bibnamefont {Timme}}, \ and\ \bibinfo
  {author} {\bibfnamefont {D.}~\bibnamefont {Witthaut}},\ }\bibfield  {title}
  {\enquote {\bibinfo {title} {Self-organized synchronization in decentralized
  power grids},}\ }\href@noop {} {\bibfield  {journal} {\bibinfo  {journal}
  {Physical Review Letters}\ }\textbf {\bibinfo {volume} {109}},\ \bibinfo
  {pages} {064101} (\bibinfo {year} {2012})}\BibitemShut {NoStop}%
\bibitem [{\citenamefont {Molnar}, \citenamefont {Nishikawa},\ and\
  \citenamefont {Motter}(2020)}]{electrical_engineering2}%
  \BibitemOpen
  \bibfield  {author} {\bibinfo {author} {\bibfnamefont {F.}~\bibnamefont
  {Molnar}}, \bibinfo {author} {\bibfnamefont {T.}~\bibnamefont {Nishikawa}}, \
  and\ \bibinfo {author} {\bibfnamefont {A.~E.}\ \bibnamefont {Motter}},\
  }\bibfield  {title} {\enquote {\bibinfo {title} {Network experiment
  demonstrates converse symmetry breaking},}\ }\href@noop {} {\bibfield
  {journal} {\bibinfo  {journal} {Nature Physics}\ }\textbf {\bibinfo {volume}
  {16}},\ \bibinfo {pages} {351--356} (\bibinfo {year} {2020})}\BibitemShut
  {NoStop}%
\bibitem [{\citenamefont {Witthaut}\ \emph {et~al.}(2022)\citenamefont
  {Witthaut}, \citenamefont {Hellmann}, \citenamefont {Kurths}, \citenamefont
  {Kettemann}, \citenamefont {Meyer-Ortmanns},\ and\ \citenamefont
  {Timme}}]{electrical_engineering3}%
  \BibitemOpen
  \bibfield  {author} {\bibinfo {author} {\bibfnamefont {D.}~\bibnamefont
  {Witthaut}}, \bibinfo {author} {\bibfnamefont {F.}~\bibnamefont {Hellmann}},
  \bibinfo {author} {\bibfnamefont {J.}~\bibnamefont {Kurths}}, \bibinfo
  {author} {\bibfnamefont {S.}~\bibnamefont {Kettemann}}, \bibinfo {author}
  {\bibfnamefont {H.}~\bibnamefont {Meyer-Ortmanns}}, \ and\ \bibinfo {author}
  {\bibfnamefont {M.}~\bibnamefont {Timme}},\ }\bibfield  {title} {\enquote
  {\bibinfo {title} {Collective nonlinear dynamics and self-organization in
  decentralized power grids},}\ }\href@noop {} {\bibfield  {journal} {\bibinfo
  {journal} {Reviews of Modern Physics}\ }\textbf {\bibinfo {volume} {94}},\
  \bibinfo {pages} {015005} (\bibinfo {year} {2022})}\BibitemShut {NoStop}%
\bibitem [{\citenamefont {Huo}\ and\ \citenamefont
  {Chen}(2023)}]{electrical_engineering4}%
  \BibitemOpen
  \bibfield  {author} {\bibinfo {author} {\bibfnamefont {L.}~\bibnamefont
  {Huo}}\ and\ \bibinfo {author} {\bibfnamefont {X.}~\bibnamefont {Chen}},\
  }\bibfield  {title} {\enquote {\bibinfo {title} {Higher-order motif-based
  time series classification for forced oscillation source location in power
  grids},}\ }\href@noop {} {\bibfield  {journal} {\bibinfo  {journal}
  {Nonlinear Dynamics}\ }\textbf {\bibinfo {volume} {111}},\ \bibinfo {pages}
  {20127--20138} (\bibinfo {year} {2023})}\BibitemShut {NoStop}%
\bibitem [{\citenamefont {Pluchino}, \citenamefont {Latora},\ and\
  \citenamefont {Rapisarda}(2005)}]{sociophysics1}%
  \BibitemOpen
  \bibfield  {author} {\bibinfo {author} {\bibfnamefont {A.}~\bibnamefont
  {Pluchino}}, \bibinfo {author} {\bibfnamefont {V.}~\bibnamefont {Latora}}, \
  and\ \bibinfo {author} {\bibfnamefont {A.}~\bibnamefont {Rapisarda}},\
  }\bibfield  {title} {\enquote {\bibinfo {title} {Changing opinions in a
  changing world: A new perspective in sociophysics},}\ }\href@noop {}
  {\bibfield  {journal} {\bibinfo  {journal} {International Journal of Modern
  Physics C}\ }\textbf {\bibinfo {volume} {16}},\ \bibinfo {pages} {515--531}
  (\bibinfo {year} {2005})}\BibitemShut {NoStop}%
\bibitem [{\citenamefont {Vasudevan}, \citenamefont {Cavers},\ and\
  \citenamefont {Ware}(2015)}]{earthquake}%
  \BibitemOpen
  \bibfield  {author} {\bibinfo {author} {\bibfnamefont {K.}~\bibnamefont
  {Vasudevan}}, \bibinfo {author} {\bibfnamefont {M.}~\bibnamefont {Cavers}}, \
  and\ \bibinfo {author} {\bibfnamefont {A.}~\bibnamefont {Ware}},\ }\bibfield
  {title} {\enquote {\bibinfo {title} {Earthquake sequencing: {C}himera states
  with {K}uramoto model dynamics on directed graphs},}\ }\href@noop {}
  {\bibfield  {journal} {\bibinfo  {journal} {Nonlinear Processes in
  Geophysics}\ }\textbf {\bibinfo {volume} {22}},\ \bibinfo {pages} {499--512}
  (\bibinfo {year} {2015})}\BibitemShut {NoStop}%
\bibitem [{\citenamefont {Strogatz}(2000)}]{reviews1}%
  \BibitemOpen
  \bibfield  {author} {\bibinfo {author} {\bibfnamefont {S.~H.}\ \bibnamefont
  {Strogatz}},\ }\bibfield  {title} {\enquote {\bibinfo {title} {From
  {K}uramoto to {C}rawford: exploring the onset of synchronization in
  populations of coupled oscillators},}\ }\href@noop {} {\bibfield  {journal}
  {\bibinfo  {journal} {Physica D: Nonlinear Phenomena}\ }\textbf {\bibinfo
  {volume} {143}},\ \bibinfo {pages} {1--20} (\bibinfo {year}
  {2000})}\BibitemShut {NoStop}%
\bibitem [{\citenamefont {Acebr{\'o}n}\ \emph {et~al.}(2005)\citenamefont
  {Acebr{\'o}n}, \citenamefont {Bonilla}, \citenamefont {Vicente},
  \citenamefont {Ritort},\ and\ \citenamefont {Spigler}}]{reviews2}%
  \BibitemOpen
  \bibfield  {author} {\bibinfo {author} {\bibfnamefont {J.~A.}\ \bibnamefont
  {Acebr{\'o}n}}, \bibinfo {author} {\bibfnamefont {L.~L.}\ \bibnamefont
  {Bonilla}}, \bibinfo {author} {\bibfnamefont {C.~J.~P.}\ \bibnamefont
  {Vicente}}, \bibinfo {author} {\bibfnamefont {F.}~\bibnamefont {Ritort}}, \
  and\ \bibinfo {author} {\bibfnamefont {R.}~\bibnamefont {Spigler}},\
  }\bibfield  {title} {\enquote {\bibinfo {title} {The {K}uramoto model: A
  simple paradigm for synchronization phenomena},}\ }\href@noop {} {\bibfield
  {journal} {\bibinfo  {journal} {Reviews of Modern Physics}\ }\textbf
  {\bibinfo {volume} {77}},\ \bibinfo {pages} {137} (\bibinfo {year}
  {2005})}\BibitemShut {NoStop}%
\bibitem [{\citenamefont {Rodrigues}\ \emph {et~al.}(2016)\citenamefont
  {Rodrigues}, \citenamefont {Peron}, \citenamefont {Ji},\ and\ \citenamefont
  {Kurths}}]{reviews4}%
  \BibitemOpen
  \bibfield  {author} {\bibinfo {author} {\bibfnamefont {F.~A.}\ \bibnamefont
  {Rodrigues}}, \bibinfo {author} {\bibfnamefont {T.~K.~D.}\ \bibnamefont
  {Peron}}, \bibinfo {author} {\bibfnamefont {P.}~\bibnamefont {Ji}}, \ and\
  \bibinfo {author} {\bibfnamefont {J.}~\bibnamefont {Kurths}},\ }\bibfield
  {title} {\enquote {\bibinfo {title} {The {K}uramoto model in complex
  networks},}\ }\href@noop {} {\bibfield  {journal} {\bibinfo  {journal}
  {Physics Reports}\ }\textbf {\bibinfo {volume} {610}},\ \bibinfo {pages}
  {1--98} (\bibinfo {year} {2016})}\BibitemShut {NoStop}%
\bibitem [{\citenamefont {Wu}\ and\ \citenamefont {Li}(2020)}]{reviews5}%
  \BibitemOpen
  \bibfield  {author} {\bibinfo {author} {\bibfnamefont {J.}~\bibnamefont
  {Wu}}\ and\ \bibinfo {author} {\bibfnamefont {X.}~\bibnamefont {Li}},\
  }\bibfield  {title} {\enquote {\bibinfo {title} {Collective synchronization
  of {K}uramoto-oscillator networks},}\ }\href@noop {} {\bibfield  {journal}
  {\bibinfo  {journal} {IEEE Circuits and Systems Magazine}\ }\textbf {\bibinfo
  {volume} {20}},\ \bibinfo {pages} {46--67} (\bibinfo {year}
  {2020})}\BibitemShut {NoStop}%
\bibitem [{\citenamefont {Motter}, \citenamefont {Zhou},\ and\ \citenamefont
  {Kurths}(2005)}]{motter2005network}%
  \BibitemOpen
  \bibfield  {author} {\bibinfo {author} {\bibfnamefont {A.~E.}\ \bibnamefont
  {Motter}}, \bibinfo {author} {\bibfnamefont {C.}~\bibnamefont {Zhou}}, \ and\
  \bibinfo {author} {\bibfnamefont {J.}~\bibnamefont {Kurths}},\ }\bibfield
  {title} {\enquote {\bibinfo {title} {Network synchronization, diffusion, and
  the paradox of heterogeneity},}\ }\href@noop {} {\bibfield  {journal}
  {\bibinfo  {journal} {Physical Review E}\ }\textbf {\bibinfo {volume} {71}},\
  \bibinfo {pages} {016116} (\bibinfo {year} {2005})}\BibitemShut {NoStop}%
\bibitem [{\citenamefont {Arenas}\ \emph {et~al.}(2008)\citenamefont {Arenas},
  \citenamefont {D{\'\i}az-Guilera}, \citenamefont {Kurths}, \citenamefont
  {Moreno},\ and\ \citenamefont {Zhou}}]{arenas2008synchronization}%
  \BibitemOpen
  \bibfield  {author} {\bibinfo {author} {\bibfnamefont {A.}~\bibnamefont
  {Arenas}}, \bibinfo {author} {\bibfnamefont {A.}~\bibnamefont
  {D{\'\i}az-Guilera}}, \bibinfo {author} {\bibfnamefont {J.}~\bibnamefont
  {Kurths}}, \bibinfo {author} {\bibfnamefont {Y.}~\bibnamefont {Moreno}}, \
  and\ \bibinfo {author} {\bibfnamefont {C.}~\bibnamefont {Zhou}},\ }\bibfield
  {title} {\enquote {\bibinfo {title} {Synchronization in complex networks},}\
  }\href@noop {} {\bibfield  {journal} {\bibinfo  {journal} {Physics Reports}\
  }\textbf {\bibinfo {volume} {469}},\ \bibinfo {pages} {93--153} (\bibinfo
  {year} {2008})}\BibitemShut {NoStop}%
\bibitem [{\citenamefont {Boccaletti}\ \emph {et~al.}(2018)\citenamefont
  {Boccaletti}, \citenamefont {Pisarchik}, \citenamefont {Del~Genio},\ and\
  \citenamefont {Amann}}]{boccaletti2018synchronization}%
  \BibitemOpen
  \bibfield  {author} {\bibinfo {author} {\bibfnamefont {S.}~\bibnamefont
  {Boccaletti}}, \bibinfo {author} {\bibfnamefont {A.~N.}\ \bibnamefont
  {Pisarchik}}, \bibinfo {author} {\bibfnamefont {C.~I.}\ \bibnamefont
  {Del~Genio}}, \ and\ \bibinfo {author} {\bibfnamefont {A.}~\bibnamefont
  {Amann}},\ }\href@noop {} {\emph {\bibinfo {title} {Synchronization: from
  coupled systems to complex networks}}}\ (\bibinfo  {publisher} {Cambridge
  University Press},\ \bibinfo {year} {2018})\BibitemShut {NoStop}%
\bibitem [{\citenamefont {Strogatz}\ and\ \citenamefont
  {Stewart}(1993)}]{strogatz1993coupled}%
  \BibitemOpen
  \bibfield  {author} {\bibinfo {author} {\bibfnamefont {S.~H.}\ \bibnamefont
  {Strogatz}}\ and\ \bibinfo {author} {\bibfnamefont {I.}~\bibnamefont
  {Stewart}},\ }\bibfield  {title} {\enquote {\bibinfo {title} {Coupled
  oscillators and biological synchronization},}\ }\href@noop {} {\bibfield
  {journal} {\bibinfo  {journal} {Scientific american}\ }\textbf {\bibinfo
  {volume} {269}},\ \bibinfo {pages} {102--109} (\bibinfo {year}
  {1993})}\BibitemShut {NoStop}%
\bibitem [{\citenamefont {Blasius}, \citenamefont {Huppert},\ and\
  \citenamefont {Stone}(1999)}]{blasius1999complex}%
  \BibitemOpen
  \bibfield  {author} {\bibinfo {author} {\bibfnamefont {B.}~\bibnamefont
  {Blasius}}, \bibinfo {author} {\bibfnamefont {A.}~\bibnamefont {Huppert}}, \
  and\ \bibinfo {author} {\bibfnamefont {L.}~\bibnamefont {Stone}},\ }\bibfield
   {title} {\enquote {\bibinfo {title} {Complex dynamics and phase
  synchronization in spatially extended ecological systems},}\ }\href@noop {}
  {\bibfield  {journal} {\bibinfo  {journal} {Nature}\ }\textbf {\bibinfo
  {volume} {399}},\ \bibinfo {pages} {354--359} (\bibinfo {year}
  {1999})}\BibitemShut {NoStop}%
\bibitem [{\citenamefont {Watts}\ and\ \citenamefont
  {Strogatz}(1998)}]{watts1998collective}%
  \BibitemOpen
  \bibfield  {author} {\bibinfo {author} {\bibfnamefont {D.~J.}\ \bibnamefont
  {Watts}}\ and\ \bibinfo {author} {\bibfnamefont {S.~H.}\ \bibnamefont
  {Strogatz}},\ }\bibfield  {title} {\enquote {\bibinfo {title} {Collective
  dynamics of ‘small-world’ networks},}\ }\href@noop {} {\bibfield
  {journal} {\bibinfo  {journal} {nature}\ }\textbf {\bibinfo {volume} {393}},\
  \bibinfo {pages} {440--442} (\bibinfo {year} {1998})}\BibitemShut {NoStop}%
\bibitem [{\citenamefont {Abrams}, \citenamefont {Pecora},\ and\ \citenamefont
  {Motter}(2016)}]{abrams2016introduction}%
  \BibitemOpen
  \bibfield  {author} {\bibinfo {author} {\bibfnamefont {D.~M.}\ \bibnamefont
  {Abrams}}, \bibinfo {author} {\bibfnamefont {L.~M.}\ \bibnamefont {Pecora}},
  \ and\ \bibinfo {author} {\bibfnamefont {A.~E.}\ \bibnamefont {Motter}},\
  }\bibfield  {title} {\enquote {\bibinfo {title} {Introduction to focus issue:
  Patterns of network synchronization},}\ }\href@noop {} {\bibfield  {journal}
  {\bibinfo  {journal} {Chaos: An Interdisciplinary Journal of Nonlinear
  Science}\ }\textbf {\bibinfo {volume} {26}} (\bibinfo {year}
  {2016})}\BibitemShut {NoStop}%
\bibitem [{\citenamefont {Thiele}\ \emph {et~al.}(2023)\citenamefont {Thiele},
  \citenamefont {Berner}, \citenamefont {Tass}, \citenamefont {Sch{\"o}ll},\
  and\ \citenamefont {Yanchuk}}]{thiele2023asymmetric}%
  \BibitemOpen
  \bibfield  {author} {\bibinfo {author} {\bibfnamefont {M.}~\bibnamefont
  {Thiele}}, \bibinfo {author} {\bibfnamefont {R.}~\bibnamefont {Berner}},
  \bibinfo {author} {\bibfnamefont {P.~A.}\ \bibnamefont {Tass}}, \bibinfo
  {author} {\bibfnamefont {E.}~\bibnamefont {Sch{\"o}ll}}, \ and\ \bibinfo
  {author} {\bibfnamefont {S.}~\bibnamefont {Yanchuk}},\ }\bibfield  {title}
  {\enquote {\bibinfo {title} {Asymmetric adaptivity induces recurrent
  synchronization in complex networks},}\ }\href@noop {} {\bibfield  {journal}
  {\bibinfo  {journal} {Chaos}\ }\textbf {\bibinfo {volume} {33}} (\bibinfo
  {year} {2023})}\BibitemShut {NoStop}%
\bibitem [{\citenamefont {Sales}, \citenamefont {Yanchuk},\ and\ \citenamefont
  {Kurths}(2024)}]{salesRecurrentChaoticClustering2024}%
  \BibitemOpen
  \bibfield  {author} {\bibinfo {author} {\bibfnamefont {M.~R.}\ \bibnamefont
  {Sales}}, \bibinfo {author} {\bibfnamefont {S.}~\bibnamefont {Yanchuk}}, \
  and\ \bibinfo {author} {\bibfnamefont {J.}~\bibnamefont {Kurths}},\ }\href
  {\doibase 10.48550/arXiv.2402.17646} {\enquote {\bibinfo {title} {Recurrent
  chaotic clustering and slow chaos in adaptive networks},}\ } (\bibinfo {year}
  {2024}),\ \Eprint {http://arxiv.org/abs/2402.17646} {arxiv:2402.17646}
  \BibitemShut {NoStop}%
\bibitem [{\citenamefont {Belykh}\ \emph {et~al.}(2008)\citenamefont {Belykh},
  \citenamefont {Osipov}, \citenamefont {Petrov}, \citenamefont {Suykens},\
  and\ \citenamefont {Vandewalle}}]{belykh2008cluster}%
  \BibitemOpen
  \bibfield  {author} {\bibinfo {author} {\bibfnamefont {V.~N.}\ \bibnamefont
  {Belykh}}, \bibinfo {author} {\bibfnamefont {G.~V.}\ \bibnamefont {Osipov}},
  \bibinfo {author} {\bibfnamefont {V.~S.}\ \bibnamefont {Petrov}}, \bibinfo
  {author} {\bibfnamefont {J.~A.}\ \bibnamefont {Suykens}}, \ and\ \bibinfo
  {author} {\bibfnamefont {J.}~\bibnamefont {Vandewalle}},\ }\bibfield  {title}
  {\enquote {\bibinfo {title} {Cluster synchronization in oscillatory
  networks},}\ }\href@noop {} {\bibfield  {journal} {\bibinfo  {journal}
  {Chaos: An Interdisciplinary Journal of Nonlinear Science}\ }\textbf
  {\bibinfo {volume} {18}} (\bibinfo {year} {2008})}\BibitemShut {NoStop}%
\bibitem [{\citenamefont {Liu}\ and\ \citenamefont
  {Chen}(2016)}]{liu2016finite}%
  \BibitemOpen
  \bibfield  {author} {\bibinfo {author} {\bibfnamefont {X.~W.}\ \bibnamefont
  {Liu}}\ and\ \bibinfo {author} {\bibfnamefont {T.~P.}\ \bibnamefont {Chen}},\
  }\bibfield  {title} {\enquote {\bibinfo {title} {Finite-time and fixed-time
  cluster synchronization with or without pinning control},}\ }\href@noop {}
  {\bibfield  {journal} {\bibinfo  {journal} {IEEE transactions on
  cybernetics}\ }\textbf {\bibinfo {volume} {48}},\ \bibinfo {pages} {240--252}
  (\bibinfo {year} {2016})}\BibitemShut {NoStop}%
\bibitem [{\citenamefont {Della~Rossa}\ \emph {et~al.}(2020)\citenamefont
  {Della~Rossa}, \citenamefont {Pecora}, \citenamefont {Blaha}, \citenamefont
  {Shirin}, \citenamefont {Klickstein},\ and\ \citenamefont
  {Sorrentino}}]{della2020symmetries}%
  \BibitemOpen
  \bibfield  {author} {\bibinfo {author} {\bibfnamefont {F.}~\bibnamefont
  {Della~Rossa}}, \bibinfo {author} {\bibfnamefont {L.}~\bibnamefont {Pecora}},
  \bibinfo {author} {\bibfnamefont {K.}~\bibnamefont {Blaha}}, \bibinfo
  {author} {\bibfnamefont {A.}~\bibnamefont {Shirin}}, \bibinfo {author}
  {\bibfnamefont {I.}~\bibnamefont {Klickstein}}, \ and\ \bibinfo {author}
  {\bibfnamefont {F.}~\bibnamefont {Sorrentino}},\ }\bibfield  {title}
  {\enquote {\bibinfo {title} {Symmetries and cluster synchronization in
  multilayer networks},}\ }\href@noop {} {\bibfield  {journal} {\bibinfo
  {journal} {Nature communications}\ }\textbf {\bibinfo {volume} {11}},\
  \bibinfo {pages} {3179} (\bibinfo {year} {2020})}\BibitemShut {NoStop}%
\bibitem [{\citenamefont {Rinzel}(2006)}]{rinzel2006bursting}%
  \BibitemOpen
  \bibfield  {author} {\bibinfo {author} {\bibfnamefont {J.}~\bibnamefont
  {Rinzel}},\ }\bibfield  {title} {\enquote {\bibinfo {title} {Bursting
  oscillations in an excitable membrane model},}\ }in\ \href@noop {} {\emph
  {\bibinfo {booktitle} {Ordinary and Partial Differential Equations:
  Proceedings of the Eighth Conference held at Dundee, Scotland, June 25--29,
  1984}}}\ (\bibinfo {organization} {Springer},\ \bibinfo {year} {2006})\ pp.\
  \bibinfo {pages} {304--316}\BibitemShut {NoStop}%
\bibitem [{\citenamefont {Izhikevich}(2000)}]{izhikevich2000neural}%
  \BibitemOpen
  \bibfield  {author} {\bibinfo {author} {\bibfnamefont {E.~M.}\ \bibnamefont
  {Izhikevich}},\ }\bibfield  {title} {\enquote {\bibinfo {title} {Neural
  excitability, spiking and bursting},}\ }\href@noop {} {\bibfield  {journal}
  {\bibinfo  {journal} {International journal of bifurcation and chaos}\
  }\textbf {\bibinfo {volume} {10}},\ \bibinfo {pages} {1171--1266} (\bibinfo
  {year} {2000})}\BibitemShut {NoStop}%
\bibitem [{\citenamefont {Han}\ and\ \citenamefont
  {Bi}(2011)}]{han2011bursting}%
  \BibitemOpen
  \bibfield  {author} {\bibinfo {author} {\bibfnamefont {X.~J.}\ \bibnamefont
  {Han}}\ and\ \bibinfo {author} {\bibfnamefont {Q.~S.}\ \bibnamefont {Bi}},\
  }\bibfield  {title} {\enquote {\bibinfo {title} {Bursting oscillations in
  {D}uffing’s equation with slowly changing external forcing},}\ }\href@noop
  {} {\bibfield  {journal} {\bibinfo  {journal} {Communications in Nonlinear
  Science and Numerical Simulation}\ }\textbf {\bibinfo {volume} {16}},\
  \bibinfo {pages} {4146--4152} (\bibinfo {year} {2011})}\BibitemShut {NoStop}%
\bibitem [{\citenamefont {Batista}\ \emph {et~al.}(2010)\citenamefont
  {Batista}, \citenamefont {Lopes}, \citenamefont {Viana},\ and\ \citenamefont
  {Batista}}]{batista2010delayed}%
  \BibitemOpen
  \bibfield  {author} {\bibinfo {author} {\bibfnamefont {C.}~\bibnamefont
  {Batista}}, \bibinfo {author} {\bibfnamefont {S.~R.}\ \bibnamefont {Lopes}},
  \bibinfo {author} {\bibfnamefont {R.~L.}\ \bibnamefont {Viana}}, \ and\
  \bibinfo {author} {\bibfnamefont {A.~M.}\ \bibnamefont {Batista}},\
  }\bibfield  {title} {\enquote {\bibinfo {title} {Delayed feedback control of
  bursting synchronization in a scale-free neuronal network},}\ }\href@noop {}
  {\bibfield  {journal} {\bibinfo  {journal} {Neural Networks}\ }\textbf
  {\bibinfo {volume} {23}},\ \bibinfo {pages} {114--124} (\bibinfo {year}
  {2010})}\BibitemShut {NoStop}%
\bibitem [{\citenamefont {Fan}\ and\ \citenamefont
  {Wang}(2017)}]{fan2017synchronization}%
  \BibitemOpen
  \bibfield  {author} {\bibinfo {author} {\bibfnamefont {D.}~\bibnamefont
  {Fan}}\ and\ \bibinfo {author} {\bibfnamefont {Q.}~\bibnamefont {Wang}},\
  }\bibfield  {title} {\enquote {\bibinfo {title} {Synchronization and bursting
  transition of the coupled {H}indmarsh-{R}ose systems with asymmetrical
  time-delays},}\ }\href@noop {} {\bibfield  {journal} {\bibinfo  {journal}
  {Science China Technological Sciences}\ }\textbf {\bibinfo {volume} {60}},\
  \bibinfo {pages} {1019--1031} (\bibinfo {year} {2017})}\BibitemShut {NoStop}%
\bibitem [{\citenamefont {Li}\ \emph {et~al.}(2023)\citenamefont {Li},
  \citenamefont {Xu}, \citenamefont {Chen}, \citenamefont {Zhou},\ and\
  \citenamefont {Yuan}}]{li2023transitions}%
  \BibitemOpen
  \bibfield  {author} {\bibinfo {author} {\bibfnamefont {R.}~\bibnamefont
  {Li}}, \bibinfo {author} {\bibfnamefont {B.~L.}\ \bibnamefont {Xu}}, \bibinfo
  {author} {\bibfnamefont {D.~B.}\ \bibnamefont {Chen}}, \bibinfo {author}
  {\bibfnamefont {J.~F.}\ \bibnamefont {Zhou}}, \ and\ \bibinfo {author}
  {\bibfnamefont {W.~J.}\ \bibnamefont {Yuan}},\ }\bibfield  {title} {\enquote
  {\bibinfo {title} {Transitions to synchronization induced by synaptic
  increasing in coupled tonic neurons with electrical synapses},}\ }\href@noop
  {} {\bibfield  {journal} {\bibinfo  {journal} {Chaos, Solitons \& Fractals}\
  }\textbf {\bibinfo {volume} {176}},\ \bibinfo {pages} {114104} (\bibinfo
  {year} {2023})}\BibitemShut {NoStop}%
\bibitem [{\citenamefont {Seliger}, \citenamefont {Young},\ and\ \citenamefont
  {Tsimring}(2002)}]{seliger2002plasticity}%
  \BibitemOpen
  \bibfield  {author} {\bibinfo {author} {\bibfnamefont {P.}~\bibnamefont
  {Seliger}}, \bibinfo {author} {\bibfnamefont {S.~C.}\ \bibnamefont {Young}},
  \ and\ \bibinfo {author} {\bibfnamefont {L.~S.}\ \bibnamefont {Tsimring}},\
  }\bibfield  {title} {\enquote {\bibinfo {title} {Plasticity and learning in a
  network of coupled phase oscillators},}\ }\href@noop {} {\bibfield  {journal}
  {\bibinfo  {journal} {Physical Review E}\ }\textbf {\bibinfo {volume} {65}},\
  \bibinfo {pages} {041906} (\bibinfo {year} {2002})}\BibitemShut {NoStop}%
\bibitem [{\citenamefont {Aoki}\ and\ \citenamefont {Aoyagi}(2009)}]{Aoki2009}%
  \BibitemOpen
  \bibfield  {author} {\bibinfo {author} {\bibfnamefont {T.}~\bibnamefont
  {Aoki}}\ and\ \bibinfo {author} {\bibfnamefont {T.}~\bibnamefont {Aoyagi}},\
  }\bibfield  {title} {\enquote {\bibinfo {title} {Co-evolution of {{Phases}}
  and {{Connection Strengths}} in a {{Network}} of {{Phase Oscillators}}},}\
  }\href {\doibase 10.1103/PhysRevLett.102.034101} {\bibfield  {journal}
  {\bibinfo  {journal} {Physical Review Letters}\ }\textbf {\bibinfo {volume}
  {102}},\ \bibinfo {pages} {034101} (\bibinfo {year} {2009})}\BibitemShut
  {NoStop}%
\bibitem [{\citenamefont {Kasatkin}\ \emph
  {et~al.}(2017{\natexlab{b}})\citenamefont {Kasatkin}, \citenamefont
  {Yanchuk}, \citenamefont {Sch{\"o}ll},\ and\ \citenamefont
  {Nekorkin}}]{kasatkin2017self}%
  \BibitemOpen
  \bibfield  {author} {\bibinfo {author} {\bibfnamefont {D.}~\bibnamefont
  {Kasatkin}}, \bibinfo {author} {\bibfnamefont {S.}~\bibnamefont {Yanchuk}},
  \bibinfo {author} {\bibfnamefont {E.}~\bibnamefont {Sch{\"o}ll}}, \ and\
  \bibinfo {author} {\bibfnamefont {V.}~\bibnamefont {Nekorkin}},\ }\bibfield
  {title} {\enquote {\bibinfo {title} {Self-organized emergence of multilayer
  structure and chimera states in dynamical networks with adaptive
  couplings},}\ }\href@noop {} {\bibfield  {journal} {\bibinfo  {journal}
  {Physical Review E}\ }\textbf {\bibinfo {volume} {96}},\ \bibinfo {pages}
  {062211} (\bibinfo {year} {2017}{\natexlab{b}})}\BibitemShut {NoStop}%
\bibitem [{\citenamefont {Kuramoto}(1984)}]{kuramoto1984chemical}%
  \BibitemOpen
  \bibfield  {author} {\bibinfo {author} {\bibfnamefont {Y.}~\bibnamefont
  {Kuramoto}},\ }\href@noop {} {\emph {\bibinfo {title} {Chemical Oscillations,
  Waves, and Turbulence}}}\ (\bibinfo  {publisher} {Springer},\ \bibinfo {year}
  {1984})\BibitemShut {NoStop}%
\bibitem [{\citenamefont {Armstrong}(1988)}]{armstrong1988groups}%
  \BibitemOpen
  \bibfield  {author} {\bibinfo {author} {\bibfnamefont {M.~A.}\ \bibnamefont
  {Armstrong}},\ }\href@noop {} {\emph {\bibinfo {title} {Groups and
  symmetry}}}\ (\bibinfo  {publisher} {Springer Verlag},\ \bibinfo {year}
  {1988})\BibitemShut {NoStop}%
\end{thebibliography}
%

\end{document}